\documentclass[3p,review,12pt]{elsarticle}

\usepackage{amssymb}
\usepackage{lineno}
\usepackage[driverfallback=dvipdfm,colorlinks,bookmarksopen,bookmarksnumbered,citecolor=blue,urlcolor=red]{hyperref}
\usepackage{float} 
\usepackage{color}
\usepackage{epstopdf}
\usepackage{amsfonts}
\usepackage{amsmath}
\usepackage{pdflscape}
\usepackage{graphicx}
\usepackage{subcaption}
\usepackage{soul}
\usepackage{xcolor}
\usepackage{bbm,bm}
\usepackage{stmaryrd}
\usepackage{color}
\usepackage{booktabs}
\usepackage{yhmath}
\usepackage{ulem} 
\usepackage{flexisym}
\usepackage{adjustbox}
\usepackage{physics}
\usepackage{boldline,multirow}
\graphicspath{{./figures/}}
\usepackage{tabularx}
\usepackage{upgreek}
\usepackage[linesnumbered,ruled,vlined]{algorithm2e}
\usepackage[labelsep=period]{caption}

\journal{Powder Technology}

\begin{document}
\begin{frontmatter}

\title{Generative modeling of granular flow on inclined planes using conditional flow matching}

\author[1]{Xuyang Li}
\author[2]{Rui Li}
\author[3]{Teng Man}
\author[2]{Yimin Lu\corref{cor}}
\ead{yimin.lu@ttu.edu}

\cortext[cor]{Corresponding author}
\address[1]{School of Construction, University of North Carolina at Charlotte, Charlotte, NC 28223, USA}
\address[2]{Department of Civil, Environmental, and Construction Engineering, Texas Tech University, Lubbock, TX 79409, USA}
\address[3]{College of Civil Engineering, Zhejiang University of Technology, Hangzhou, Zhejiang 310023, China}

\begin{abstract}
Granular flows govern many natural and industrial processes, yet their interior kinematics and mechanics remain largely unobservable, as experiments access only boundaries or free surfaces. Conventional numerical simulations are computationally expensive for fast inverse reconstruction, and deterministic models tend to collapse to over-smoothed mean predictions in ill-posed settings. This study, to the best of the authors' knowledge, presents the first conditional flow matching (CFM) framework for granular-flow reconstruction from sparse boundary observations. Trained on high-fidelity particle-resolved discrete element simulations, the generative model is guided at inference by a differentiable forward operator and a novel sparsity-aware gradient guidance mechanism. This mechanism avoids the gradient dilution inherent to standard mean-squared-error approaches, preserves the absolute physical scale of observation errors, enforces measurement consistency without hyperparameter tuning, and prevents unphysical velocity predictions in non-material regions. A physics decoder maps the reconstructed velocity fields to stress states and energy fluctuation quantities, including mean stress, deviatoric stress, and granular temperature. The framework accurately recovers interior flow fields from full observation to only 16\% of the informative window, and it remains effective under strongly diluted spatial resolution with only 11\% of data. It also outperforms a deterministic CNN baseline in the most ill-posed reconstruction regime and provides spatially resolved uncertainty estimates through ensemble generation. These results demonstrate that conditional generative modeling offers a practical route for non-invasive inference of hidden bulk mechanics in granular media, and it suggests potential applicability for inverse problems in particulate and multiphase systems.
\end{abstract}

\begin{keyword}
Granular mechanics \sep Diffusion model \sep Generative AI \sep Training-free inference \sep Uncertainty quantification \sep Inverse problem
\end{keyword}

\end{frontmatter}

\section{Introduction}
\label{sec_intro}
Granular materials are ubiquitous in geophysical and industrial systems \cite{jaeger1996granular}, and their flows govern various processes, from landslides and avalanches \cite{buscarnera2013soil,lu2019centrifuge,soga2016trends} to the transport and handling of pharmaceutical powder, milled biomass, and ores \cite{fan2017segregation,lu2022wedge,lu2023effects}. Despite rapid advancements in experimental techniques, investigating the fundamental physics of granular flows remains severely restricted by limited observability. For example, in canonical setups like the inclined plane flow experiments, while advanced optical imaging and particle tracking techniques can capture particle kinematics at boundaries and the free surface, the opaque bulk makes the interior velocity, stress, and fluctuation fields practically unobservable \cite{wu2025unified,jop2005crucial,lueptow2000piv,zhou2025rate,gollin2017performance}. Knowing these internal states is physically critical because the macroscopic behavior and catastrophic failure of granular media are dictated by hidden micro-mechanisms, such as localized shear banding, force chain transmission, and internal energy dissipation, which may not be completely recoverable solely from exterior surface manifestations \cite{man2025run,wang2023micro,lu2024shear,man2025grain}. Without knowing the interior mechanical and flow behavior, our fundamental understanding of granular rheology remains incomplete.

To probe these hidden states, researchers traditionally rely on numerical simulations. The discrete element method (DEM) \cite{cundall1979discrete,zhu2007discrete} can resolve particle-scale contacts and can recover hidden bulk kinematics and stresses. However, it is computationally prohibitive for large-scale systems, and its predictive accuracy depends strongly on contact-law choices and parameter calibration \cite{deak2025high,xia2020review,jin2022fidelity,li2025data}, which poses uncertainty since various DEM parameters often need to be calibrated from sparse macroscopic observations. Alternatively, continuum mechanics-based approaches, such as the Finite Element Method (FEM), Material Point Method (MPM), and Smoothed Particle Hydrodynamics (SPH), offer greater macroscopic efficiency \cite{zheng2015finite,dunatunga2015continuum,lu2021flow,hurley2017continuum,mu2026soil,kumar2017modelling}. However, these continuum methods require constitutive assumptions and numerical treatments \cite{xue2023nonlocal}, making them struggle to universally capture the complex, transitional nature of granular flows (e.g., free surfaces, flow-jamming transition, and creeping) \cite{lu2021arch}. Crucially, all these numerical techniques are inherently forward modeling approaches. They strictly require fully defined initial conditions, boundary conditions, and material properties to execute, lacking the flexibility to directly infer full-field mechanics from sparse experimental observations without relying on computationally expensive, iterative trial-and-error calibration.

This limitation has motivated data-driven reconstruction methods. Linear Stochastic Estimation (LSE)~\cite{baars2016spectral} and \textcolor{black}{Gappy Proper Orthogonal Decomposition (Gappy POD)~\cite{everson1995karhunen, willcox2006unsteady}, commonly used as lightweight baselines for sparse flow reconstruction in fluid mechanics, can interpolate missing data by projecting} sparse observations onto dominant modes. \textcolor{black}{However,} their linearity fundamentally fails to capture the \textcolor{black}{highly} nonlinear \textcolor{black}{and chaotic} dynamics of granular media.
Deep learning techniques, including Convolutional Neural Networks (CNNs)~\cite{xu2022improved, lu2021machine, bolandi2022bridging}, can learn richer nonlinear mappings and have become increasingly common in mechanics and granular-material modeling. 
Advanced architectures like Physics-Informed Neural Networks (PINNs)~\cite{raissi2019physics, bolandi2023physics} and Neural Operators~\cite{lu2021learning, li2020fourier} further broaden this landscape. However, while PINNs often struggle in this domain due to the lack of closed-form partial differential equations (PDEs) capable of fully describing complex granular rheology, Neural Operators can learn continuous functional mappings without relying on explicit PDEs, yet they are inherently restricted to producing a single deterministic output field.
More fundamentally, standard deep learning architectures act as deterministic mapping. When applied to ill-posed inverse problems, such as mapping sparse boundary data to full interior fields, \textcolor{black}{deterministic models trained with pointwise regression losses tend to produce conditional-mean predictions \cite{bishop2006pattern}}. This results in catastrophic over-smoothing that strips away critical fine-scale physical fluctuations (e.g., localized granular temperature and stochasticity caused by different particle sampling and packing) and fails to quantify the predictive epistemic uncertainty inherent to incomplete observational data.

Generative modeling offers a more suitable probabilistic paradigm because it aims to learn not just one reconstruction, but the conditional distribution of admissible interior states. While early architectures like Variational Autoencoders (VAEs)~\cite{kingma2013auto} and Generative Adversarial Networks (GANs)~\cite{goodfellow2014generative} suffer from blurry outputs and training instability, Diffusion Models~\cite{ho2020denoising, song2020score} achieve high fidelity; \textcolor{black}{but they still suffer from higher iterative sampling costs because their curved stochastic trajectories necessitate hundreds of small numerical integration steps}. Recently, Flow Matching (FM) \cite{tong2023improving, lipman2022flow} has emerged as a highly efficient alternative, learning deterministic vector fields to transport probability density using continuous ordinary differential equations (ODEs). \textcolor{black}{By constructing straight-path trajectories, FM allows for significantly larger integration steps, thereby drastically reducing the number of network evaluations during inference compared to standard diffusion models \cite{lipman2022flow, dasgupta2026solving}.}
In particular, the recent conditional flow matching has been pioneered with great success to reconstruct near-wall fluid turbulence and quantify uncertainty from sparse measurements \cite{parikh2025conditional}. \textcolor{black}{More broadly, the paradigm of guiding pre-trained generative priors with differentiable forward models to solve physics-constrained inverse problems is rapidly advancing across various scientific domains, such as real-time flood mapping \cite{wu2025piff}.} 

However, despite its success in continuous fluids, the application of flow matching to granular mechanics remains unexplored due to fundamental physical hurdles. Granular flows present distinct challenges compared to space-filling continuous fluids: they are characterized by extreme spatial sparsity, dynamically changing void (non-material and zero-velocity) regions, and highly chaotic particle-level interactions. Standard continuous generative models inherently struggle to preserve these abrupt spatial discontinuities. Therefore, a rigorous generative framework is greatly needed to address this gap, one capable of inferring hidden granular physics while systematically handling spatial intermittency, quantifying predictive uncertainty, and avoiding the prohibitive computational costs of iterative sampling.

In this work, we propose a novel generative learning pipeline based on conditional flow matching~\cite{lipman2022flow, tong2023improving, parikh2025conditional} to reconstruct internal granular kinematics and mechanics from sparse boundary velocity observations, which aligns with what can be obtained from state-of-the-art granular flow experiments. Our framework couples a generative backbone with a differentiable forward operator that maps interior states to observable boundary signals, so that sampling can be guided toward boundary-consistent bulk realizations without retraining. To explicitly accommodate the extreme spatial sparsity and discrete nature of granular media, a sparsity-aware gradient guidance mechanism is introduced, which intrinsically preserves the absolute physical scale of observation errors, eliminating the need for empirical hyperparameter tuning while preventing unphysical numerical predictions in non-material regions.
It is demonstrated that this approach not only faithfully reconstructs high-fidelity instantaneous velocity fields and rheological descriptors, such as stress states and granular temperature, but also provides inherent uncertainty quantification~\cite{psaros2023uncertainty} without requiring model retraining. 
Ultimately, this fast, training-free paradigm resolves the ill-posed nature of granular inverse reconstruction, enabling non-invasive physical discovery, and it \textcolor{black}{provides a possible route for future digital-twin applications} in complex multiphase systems.

\section{Methodology}
\subsection{Problem statement}
This paper focuses on gravity-driven dense granular flow down inclined planes and seek to reconstruct interior flow states from limited observations available near the boundary. In this study, the target interior quantities include the velocity field, velocity fluctuation (i.e., granular temperature), and mechanical fields (i.e., stresses). The objective is to infer the conditional distribution of valid interior states given boundary velocity observations, rather than a single deterministic estimate, because multiple interior realizations may remain consistent with sparse measurements. 
This problem setting follows directly from the limited observability of granular flows in experiments, where sidewall or free-surface measurements are often available, whereas the bulk kinematics and stress transmission within the flowing layer are much more difficult to access directly. Formulated as a highly ill-posed inverse problem, the developed framework seeks the conditional generation without the need for computationally expensive and iterative simulations.

The feasibility of this inverse problem is fundamentally grounded in the continuum mechanics and non-local rheology of dense granular media. Macroscopically, in inclined dense granular flows, the velocity profile and stress distribution are governed by depthwise momentum balance together with the material rheology, so the near-boundary kinematics necessarily reflect the integrated response of the internal flowing layer. Previous studies show that internal shear stress, normal stress, and velocity gradients are coupled through constitutive structure rather than evolving independently \cite{jop2006constitutive}. In addition, boundary conditions (e.g., basal and side-wall friction) modify the entire flow profile, which further indicates that measurable boundary behavior carries information about the interior state. 

Microscopically, dense granular materials exhibit strong non-local rheological behaviors that couple boundary layers to interior \cite{kamrin2015nonlocal}. At the grain scale, force transmission occurs through spatially extended contact networks, so stresses and fluctuations near the boundary can remain correlated with structures deeper in the bulk, although those correlations weaken with distance. These physical couplings motivate learning a conditional mapping from boundary observations to interior kinematic realizations, and then inferring stress quantities from the reconstructed velocity field.

\textcolor{black}{The boundary-to-interior correlation should not be interpreted as a deterministic one-to-one mapping over an unlimited distance. Instead, it is expected to weaken with the distance from the observed boundary. In the present study, the three reconstructed slices are located at $y^+ = 6.7d$, $13.3d$, and $20d$, respectively, where $d$ is the mean particle diameter (Figure~\ref{fig:dem}(a)). This range is comparable to the mesoscale length over which dense granular flows exhibit nonlocal and boundary-influenced behavior. Previous studies have shown that velocity fluctuation, force-chain correlation, and side-wall impacts can persist from several to tens of grain diameters \cite{jop2005crucial,pouliquen2004velocity,majmudar2005contact}. Therefore, the deepest slice is not assumed to be fully determined by the boundary observation, but it is expected to remain partially constrained by the coupled kinematics, stress transmission, and nonlocal rheology of the dense flow, making it important to consider such inference uncertainty.}

\subsection{Numerical model setup and dataset construction} 
The dataset used in this study is generated from direct numerical simulations of spherical particles flowing down inclined planes using DEM \cite{man2025run,man2026propagation}, \textcolor{black}{which has been validated in previous research \cite{man2021finite,wang2025collapse}. In DEM simulations, we used spherical particles (particle density $\rho_p = 2.65$ g/cm$^3$) with a Hookean linear spring-dashpot contact model. Detailed information of governing equations, normal contact forces, and tangential contact forces were omitted here and can be found in the previous work \cite{man2025run}. The normal and tangential stiffness of particles were $k_n = 1\times 10^5$ N/m and $k_t = 4\times 10^4$ N/m, which ensured that particles only exhibit small deformations. The restitution coefficient is $e = 0.8$, so that the normal damping coefficient can be calculated based on $e = \exp[-\pi\gamma_n/(2\sqrt{k_n/m_e - (\gamma_n/2)^2})]$, where $\gamma_n$ is the damping coefficient and $m_e = (m_i^{-1} + m_j^{-1})^{-1}$ is the reduced mass of a contact pair. In order to make the simulation stable, we set up the integration time step $\Delta t = 1\times 10^{-6}$ s, based on $\Delta t \leq 0.1\sqrt{m/k_n}$, and we sampled the result files every 0.001 s.} 

\begin{figure}[h!]
    \centering
    \includegraphics[width=\linewidth]{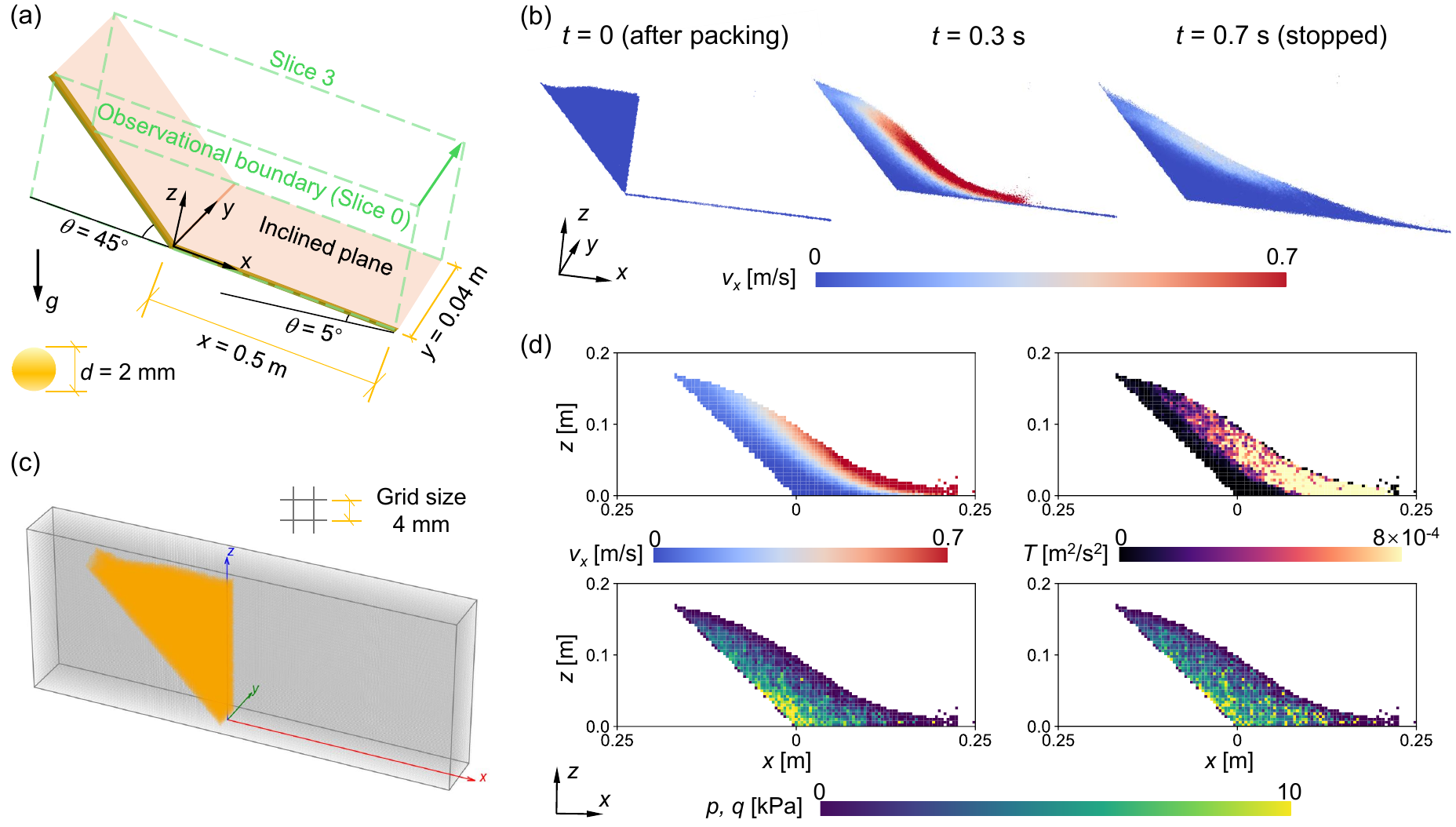}
    \caption{Numerical modeling and dataset construction. (a) Schematic view of the discrete element method (DEM) model showing the geometry, particle size, coordinate system, and representative surfaces of interest, from the observational boundary (Slice 0) to the other boundary plane (Slice 3). (b) Representative DEM results, showing runout velocity $v_x$ profile after packing, during the flow, and after flow stops. (c) Demonstration of the grid-averaging scheme that transfers particle-scale information into \textcolor{black}{grid cells}, with a grid size of 4~mm. (d) Demonstrative grid-averaged results at $t=0.3$~s used for conditional flow matching, including runout velocity $v_x$, granular temperature $T$, mean stress $p$, and deviatoric stress $q$.}
    \label{fig:dem}
\end{figure}

Figure \ref{fig:dem}(a) presents the setup of the DEM model, where the system consists of an initially static granular packing and four parts of boundary conditions: (1) inclined plane \#1 with a 5$^{\circ}$ inclination angle; (2) inclined plane \#2, which has an inclination angle of 50$^{\circ}$ and connects with the plane \#1; (3) two lateral walls (not shown in the figure) that prevent particles to escape laterally from the collapsing channel; and (4) another unseen plane, i.e., a temporary retaining plate to hold the material during particle initialization, which passes through the intersection line between planes \#1 and \#2 and is perpendicular to plane \#1. Note that on the left side ($-x$ direction) of the retaining plate, an initial granular packing is created with approximately $95,000$  particles. The particle diameter follows a uniform distribution that varies from $0.8d$ to $1.2d$, where the average diameter, $d$, is 2.0~mm. The retaining wall is used to hold the particles during initial granular packing and keep them static before the simulation starts (shown as the $t=0$ frame in Figure \ref{fig:dem}(b)). The particle-particle and particle-wall friction coefficients are 0.2 and 0.5, respectively. At $t = 0$, the retaining plate is removed to release the granular system to flow along the channel and eventually deposit on the plane \#1 (Figure \ref{fig:dem}(b)). It should be noted that plane \#1 in this research is covered by a layer of uniformly fixed spherical particles with a diameter of 4 mm, which generates a rough and rigid surface.

While DEM simulations provide exhaustive micro-scale information (e.g., individual particle trajectories, force chain evolution), key granular mechanical behavior (e.g., shear banding, energy dissipation) closely relates to \textcolor{black}{continuum-mechanics-based }physical properties, such as stress tensor $\bm{\sigma}$ and granular temperature $T$. Accordingly, the particle-scale data is projected onto a fixed Eulerian grid and is used to compute locally averaged kinematic and mechanical quantities for each cell (Figure~\ref{fig:dem}(c)). This representation provides a physically interpretable bridge between discrete grain dynamics and continuum-scale flow behavior, while defining a consistent state space for probabilistic reconstruction.

For each cell, the velocity components are simply averaged from $n$ particles whose center fall into the cell, i.e., $\bm{v}_{cell} = \frac{1}{n} \sum \bm{v}_i$, where $i$ stands for each particle that falls in a specific cell. In each cell, the granular temperature $T$ is calculated as \cite{fan2011shear}:
\begin{equation}
    T = \frac{\left(\overline{v_x'v_x'} + \overline{v_y'v_y'} + \overline{v_z'v_z'}\right)}{3}
\end{equation}
where the overline $\overline{(\cdot)}$ denotes the average over all particles within the cell. $v_x'$, $v_y'$, $v_z'$ are components of the velocity fluctuation vector $\bm{v}'_i$ for each particle $i$ within the cell, defined as $\bm{v}'_i = \bm{v}_i - \bm{v}_{cell}$.

The stress tensor $\bm{\sigma}$ for each cell is calculated according to François Nicot et al. \cite{nicot2013definition} as the sum of a static term given by the Love--Weber formula related to contact forces, and a dynamic term accounting for inertial effects related to rotation velocities and accelerations of the particles. \textcolor{black}{As the velocity in our simulation is relatively small and the inertial effects are mostly negligible, only the static contact force contribution is considered in this study}:
\begin{equation}
\langle\sigma_{ij}\rangle = \frac{1}{V}\sum_{c=1}^{n_c} {f}_i^c {l}_j^c
\end{equation}
where $n_c$ is the number of contacts within the cell volume $V$, ${f}_i^c$ is the $i$th component of the contact force at contact $c$, ${l}_j^c$ is the $j$th component of the branch vector connecting the centers of the two particles at contact $c$. 

In this work, the mean stress $p$ and deviatoric stress $q$ are utilized to represent the full stress tensor $\bm{\sigma}$: 

\begin{equation}
p = -\frac{1}{3}\left(\sigma_{xx} + \sigma_{yy} + \sigma_{zz}\right)
\end{equation}

\begin{equation}
q = \sqrt{\frac{1}{2}\left[\left(\sigma_{xx} 
- \sigma_{yy}\right)^2 + \left(\sigma_{yy} 
- \sigma_{zz}\right)^2 + \left(\sigma_{xx} 
- \sigma_{zz}\right)^2 + 6\left(\sigma_{xy}^2 
+ \sigma_{xz}^2 + \sigma_{yz}^2\right)\right]}
\end{equation}
where $\sigma_{xx}$, $\sigma_{yy}$, and $\sigma_{zz}$ are normal stress components, and $\sigma_{xy}$, $\sigma_{xz}$, and $\sigma_{yz}$ are shear stress components of the stress tensor. The mean stress $p$ measures the overall pressure on \textcolor{black}{each cell} that results in volume change, while the deviatoric stress $q$ characterizes the magnitude of shear stress that triggers shape change within the granular assembly.

A typical demonstration of grid-based velocity ($v_x$), granular temperature $T$, and stresses $p$ and $q$ during the flow is shown in Figure~\ref{fig:dem}(d).

\subsection{Generative modeling framework based on conditional flow matching}
\begin{figure}[H]
\centering
\includegraphics[width=\textwidth]{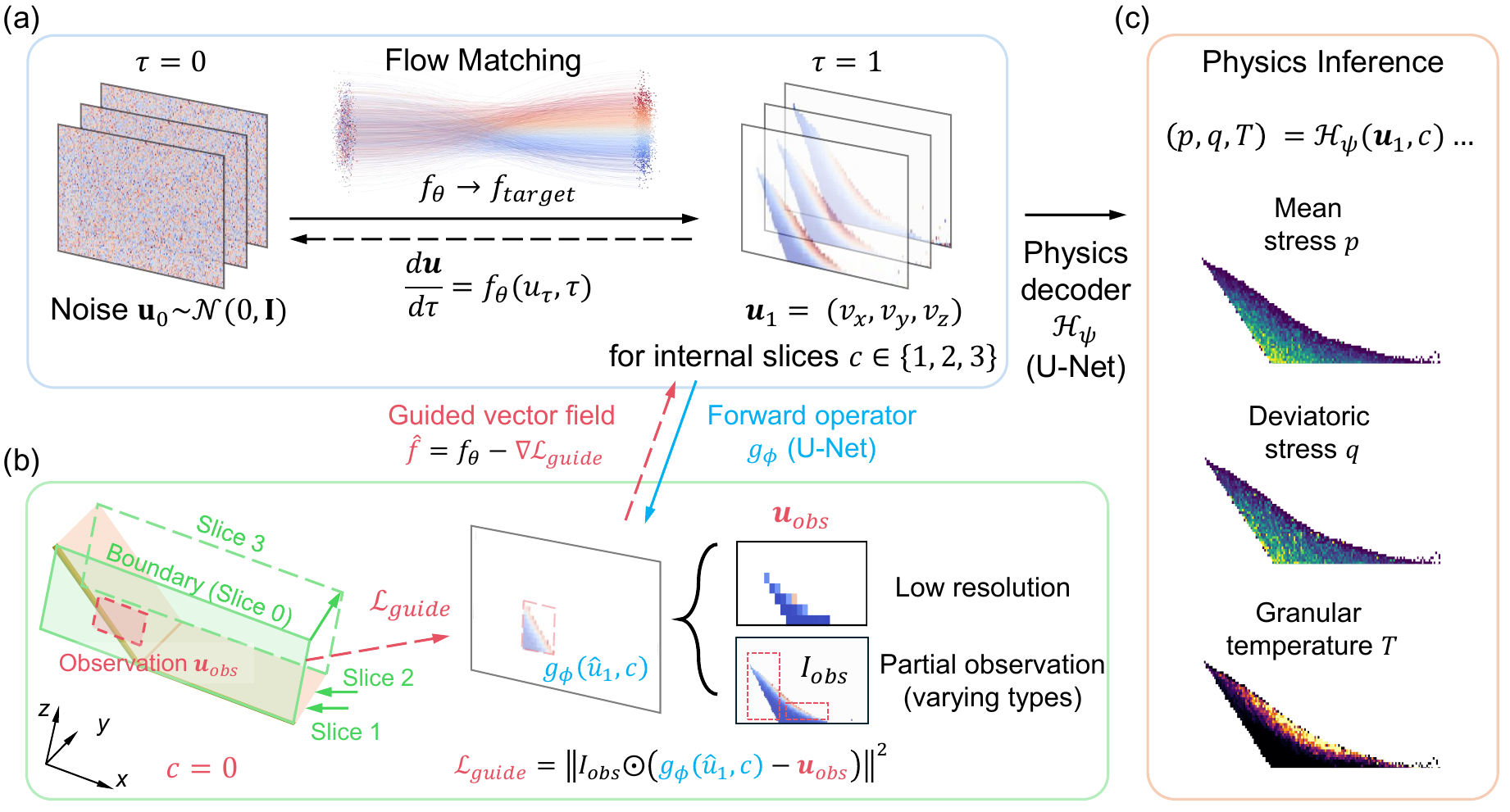}
\caption{Overview of the proposed framework. The architecture integrates (a) a continuous-time probabilistic flow matching backbone as a kinematic prior $f$, (b) a differentiable neural forward operator $g$ with a gradient guidance mechanism for conditional flow matching to infer internal particle dynamics from partial boundary observations, and (c) a physics decoder $\mathcal{H}$ to infer key physical properties from the reconstructed kinematic fields.}
\label{fig:framework}
\end{figure}

The proposed generative framework leverages CFM to reconstruct unobservable internal granular kinematics and thermodynamic states from sparse boundary measurements. As illustrated in Figure~\ref{fig:framework}(a-b), the architecture seamlessly integrates a continuous-time probabilistic backbone $f$ acting as a kinematic prior, a differentiable neural forward operator $g$, and a gradient guidance mechanism. Furthermore, Figure~\ref{fig:framework}(c) illustrates a physics decoder $\mathcal{H}$ designed to infer key thermodynamic and mechanical properties directly from the reconstructed velocity fields. Together, these components enable robust end-to-end inference of complex internal flow fields alongside explicit uncertainty quantification.

\subsubsection{Probabilistic generative backbone}
The core of our framework is a continuous-time generative model designed to approximate the intrinsic data distribution $P(\mathbf{u})$ of the interior velocity fields $\mathbf{u}$.
To prevent semantic ambiguity, it is important to note that the state variable $\mathbf{u}$ represents the actual physical kinematic velocity of the granular media (i.e., $\mathbf{u}=\bm{v}_{cell}$), comprising the three-dimensional spatial components ($v_x, v_y,$ and $v_z$) evaluated at each respective slice. Conversely, the abstract vector fields associated with the flow matching process (e.g., $f_\tau$) refer exclusively to the mathematical transport velocity driving the probability density evolution across the virtual time domain. 
Crucially, this generative framework is trained to function as a generic kinematic prior: it maps unstructured noise to physically valid granular flow states, independent of any specific observations, by parameterizing the mathematical transport vector field ($f_\tau$).

Let $\mathbf{u}_0$ denote a sample from the standard Gaussian prior distribution $P_0(\mathbf{u}_0) = \mathcal{N}(\mathbf{0}, \mathbf{I})$, representing the initial unstructured noise. Let $\mathbf{u}_1$ denote a sample from the target data distribution $Q(\mathbf{u}_1)$, representing the physical velocity fields obtained from DEM simulations. 
Here, the target distribution $Q$ is empirically defined by the collection of these high-fidelity simulation samples.
FM defines a time-dependent probability density path $P_{\tau}$ for virtual time horizon $\tau \in [0, 1]$, which smoothly interpolates between the noise distribution $P_0$ and the target data distribution $Q$.

Unlike diffusion models that rely on stochastic sampling steps, the evolution of samples in FM is governed by a deterministic ordinary differential equation (ODE):
\begin{equation}
\frac{d\mathbf{u}_\tau}{d\tau} = f_\tau(\mathbf{u}_\tau), \quad \mathbf{u}_{\tau=0} = \mathbf{u}_0
\end{equation}
where $\mathbf{u}_\tau$ represents the intermediate state of the velocity field sample at virtual time $\tau$, evolving from the noise sample $\mathbf{u}_0$ ($\mathbf{u}_0 \sim P_0$) to the data sample $\mathbf{u}_1$.
The term $f_\tau(\cdot)$ denotes the time-dependent vector field driving this transport.

Conceptually, integrating this ODE transports a noise sample $\mathbf{u}_0$ along a continuous trajectory to form a realistic velocity field $\mathbf{u}_1$.
Although the ODE acts on individual samples, the evolution of all such samples drawn from $\mathbf{u}_0 \sim P_0$ transports the entire probability mass from the noise distribution to the data distribution.
Our objective is to parameterize this vector field using a neural network $f_\theta(\mathbf{u}, \tau, c)$ (implemented as a U-Net and detailed in Section \ref{sec:training}), such that $f_\theta \approx f_\tau$. Here, $c$ is introduced as a conditioning variable representing the physical depth (slice index) of the target evaluation plane (Figure~\ref{fig:framework}(b)).
By training on a vast ensemble of such noise-to-data trajectories, the network implicitly learns the underlying vector field of the entire population. 
Consequently, during inference, the model can generate realistic flow fields (which were not seen during training) by integrating the ODE from $\tau=0$ to $\tau=1$ starting from sampled noise.

For the specific choice of trajectory, the interpolation path is defined using Optimal Transport (OT) \cite{lipman2022flow}, which corresponds to a straight-line trajectory between a noise sample $\mathbf{u}_0$ and a target data sample $\mathbf{u}_1$.
The intermediate kinematic state $\mathbf{u}_\tau$ at any virtual time $\tau$ is constructed via linear interpolation:
\begin{equation}
\mathbf{u}_\tau = (1 - (1 - \sigma_{\min})\tau)\mathbf{u}_0 + \tau \mathbf{u}_1
\label{eq:ot_path}
\end{equation}
where $\sigma_{\min}$ is a small constant (set to $10^{-4}$) introduced to ensure numerical stability near $\tau=1$. $\mathbf{u}_\tau$ is the interpolated current state.

Taking the time derivative of Eq.~\eqref{eq:ot_path} yields the target vector field $f_{target}$ required for training:
\begin{equation}
f_{target}=\frac{d\mathbf{u}_\tau}{d\tau} = \mathbf{u}_1 - (1 - \sigma_{\min})\mathbf{u}_0
\label{eq:theory_target}
\end{equation}

However, in the practical implementation, it is numerically advantageous to express the target vector field in terms of the current state $\mathbf{u}_\tau$ and the target data $\mathbf{u}_1$, rather than the initial noise $\mathbf{u}_0$. By solving Eq.~\eqref{eq:ot_path} for $\mathbf{u}_0$ and substituting it into Eq.~\eqref{eq:theory_target}, we obtain the equivalent form used in training:
\begin{equation}
f_{target}(\tau,\mathbf{u}_\tau | \mathbf{u}_1) = \frac{\mathbf{u}_1 - (1 - \sigma_{\min})\mathbf{u}_\tau}{1 - (1 - \sigma_{\min})\tau}
\label{eq:implementation_target}
\end{equation}
This formulation ensures that the target vector field is computed consistently with the network inputs in the computational graph.

The network parameters $\theta$ are optimized by minimizing the CFM objective:
\begin{equation}
\mathcal{L}_{CFM}(\theta) = \mathbb{E}_{\tau\in [0, 1], c \sim \{1,2,3\}, \mathbf{u}_1 \sim Q(\cdot|c), \mathbf{u}_0 \sim P_0} \left[ \left\| f_\theta(\mathbf{u}_\tau, \tau, c) - f_{target}(\tau,\mathbf{u}_\tau | \mathbf{u}_1)\right\|^2_2 \right]
\label{eq:loss_general}
\end{equation}
where $\mathbb{E}$ denotes the mathematical expectation taken over the uniformly sampled virtual time $\tau \in [0, 1]$, the target physical states $\mathbf{u}_1 \sim Q$, and the initial Gaussian noise $\mathbf{u}_0 \sim P_0$.

\subsubsection{Forward operator as a differentiable surrogate}\label{sec:forward_operator}
To enable guided inference, a differentiable mechanism is required to map the generated interior velocity fields to the observable boundary velocity fields. In granular flows, this mapping is governed by particle-particle contacts (i.e., collision and friction), which are conventionally resolved using computationally intensive DEM simulations. Since running DEM iteratively during the reverse generative process is intractable, a deterministic neural surrogate $g_\phi$ is trained to efficiently approximate this mechanical relationship.
Specifically, a deterministic formulation is adopted under the premise that the reconstruction uncertainty arises primarily from the ill-posed nature of the inverse problem (limited boundary observability) rather than the modeling error.

The surrogate $g_\phi$ is parameterized using a U-Net with learnable parameters $\phi$, conditioned on the physical layer index. Given an interior velocity slice $\mathbf{u}_{in}$ and a layer indicator $c \in \{1, 2, 3\}$ (corresponding to physical locations of $y^+ = 6.7d$, $13.3d$, and $20d$, where $d$ is particle diameter, and $y^+ = 20d$ represents the non-observational boundary surface, see Figure~\ref{fig:framework}(b)), the network predicts the corresponding boundary velocity $\hat{\mathbf{u}}_{obs}$ (at slice 0):
\begin{equation}
\hat{\mathbf{u}}_{obs} = g_\phi(\mathbf{u}_{in}, c)\label{eq:fwd}
\end{equation}
Here, the conditioning variable $c$ is projected into a learnable embedding vector and added to the feature maps within the U-Net residual blocks, explicitly encoding the physical depth into the layer-wise feature.
 
The network is trained in a supervised manner using paired samples from the DEM simulations. Optimization is performed by minimizing the Mean Squared Error (MSE) between the predicted and ground-truth boundary velocities, denoted as the forward modeling loss:
\begin{equation}
\mathcal{L}_{fwd}(\phi) = \mathbb{E}_{(\mathbf{u}_{in}, \mathbf{u}_{obs}, c) \sim \mathcal{D}} \left[ \left\| g_\phi(\mathbf{u}_{in}, c) - \mathbf{u}_{obs} \right\|^2_2 \right]
\end{equation}
Once trained, the parameters $\phi$ are frozen. By efficiently approximating the forward mechanical response, $g_\phi$ serves as a differentiable surrogate during the inference phase, providing gradient signals to enforce consistency with boundary measurements.

\subsubsection{Gradient-guided inference}
\label{sec:gradient-guided-inference}
With the pre-trained generative backbone $f_\theta$ and the differentiable operator $g_\phi$, we can now formulate the reconstruction task as a guided generation process.
During standard unguided inference, integrating the learned vector field $f_\theta$ from a Gaussian prior $\mathbf{u}_0 \sim \mathcal{N}(\mathbf{0}, \mathbf{I})$ generates an unconditional velocity sample. However, to solve the inverse problem, this generative trajectory must be dynamically adjusted to satisfy the specific boundary measurements $\mathbf{u}_{obs}$ of a given test instance.

At any intermediate virtual time $\tau$, we leverage the straight-path property of optimal transport flow matching to estimate the expected terminal state $\hat{\mathbf{u}}_1$ directly from the current state $\mathbf{u}_\tau$:
\begin{equation}
\hat{\mathbf{u}}_1 = \mathbf{u}_\tau + (1 - \tau) f_\theta(\mathbf{u}_\tau, \tau, c)
\end{equation}

This estimated state $\hat{\mathbf{u}}_1$ is then projected through the forward operator $g_\phi$ using Eq.~\ref{eq:fwd} to predict boundary velocity $\hat{\mathbf{u}}_{obs}$ (at slice 0).
The guidance loss $\mathcal{L}_{guide}$ is thus formulated as a Sum-of-Squared-Errors (SSE):
\begin{equation}
\mathcal{L}_{guide}(\mathbf{u}_\tau) = \left\| \mathbf{I}_{obs} \odot \left( g_\phi(\hat{\mathbf{u}}_1, c) - \mathbf{u}_{obs} \right) \right\|_2^2 \label{eq:cons_loss}
\end{equation}
where $\odot$ denotes the element-wise product, and $\| \cdot \|_2^2$ represents the squared $L_2$ norm (sum of squared elements). 

The choice of the unnormalized SSE, rather than the standard Mean Squared Error (MSE), is a deliberate design to accommodate the spatial sparsity inherent to granular flows. 
Unlike continuous fluids, granular systems feature extensive non-material, zero-velocity space. A standard MSE formulation would artificially dilute the correction gradient $\nabla_{\mathbf{u}_\tau} \mathcal{L}_{guide}$ by averaging across these empty grids. The spatially selective SSE prevents this dilution, ensuring the gradient preserves its proper physical magnitude. This formulation is hereafter referred to as the sparsity-aware gradient guidance.

To compute this correction, the instantaneous gradient $\nabla_{\mathbf{u}_\tau} \mathcal{L}_{guide}$ is evaluated at each time step $\tau$ via full backpropagation through both the forward mapping $g_\phi$ and the generative network $f_\theta$. This ensures that the adjustments are strictly regularized by the measurement at each time step.

\textcolor{black}{It is important to acknowledge the physical implications of this learned forward mapping. In actual granular flows, the interior velocity field $\mathbf{u}_{in}$ and the boundary observation $\mathbf{u}_{obs}$ are simultaneously determined by the full 3D discrete particle dynamics; there is no causal relationship where the interior state physically produces the boundary state. The surrogate $g_\phi$ therefore learns a statistical correlation rather than a causal forward process. Consequently, backpropagating the gradient $\nabla_{\mathbf{u}_\tau} \mathcal{L}_{guide}$ during the guided inference operates under the implicit assumption that the neural network surrogate adequately captures the gradient of the true physical manifold, even when evaluated at intermediate, potentially non-physical states $\hat{\mathbf{u}}_1$. While this approximation remains a statistical proxy for the underlying complex physical correlations, it is pragmatically effective for guiding the generation toward measurement consistency.}

The guided vector field $\tilde{f}_\tau$ is given by: 
\begin{equation}
\tilde{f}_\tau(\mathbf{u}_\tau, \tau, c) = f_\theta(\mathbf{u}_\tau, \tau, c) - \xi \nabla_{\mathbf{u}_\tau} \mathcal{L}_{guide}(\mathbf{u}_\tau)
\label{eq:guided_velocity}
\end{equation}
where $\xi > 0$ is the guidance scale. Crucially, because the SSE preserves the absolute physical magnitude of the discrepancy, the resulting gradient naturally scales. Empirically, this formulation mitigates hyperparameter sensitivity, allowing $\xi$ to be robustly fixed at $1$ across all test instances.

Numerically, the guided probability flow ODE is solved using a first-order Euler discretization ($\Delta \tau = 1/N$). At step $i$, the state is updated by:
\begin{equation}
\mathbf{u}_{\tau_{i+1}} = \mathbf{u}_{\tau_i} + \tilde{f}_{\tau_i}(\mathbf{u}_{\tau_i}, \tau_i, c) \cdot \Delta \tau
\end{equation}
This iterative integration gradually refines the initial unstructured noise $\mathbf{u}_0$ into a high-fidelity kinematic state $\mathbf{u}_1$ that satisfies the sparse boundary constraints.

\subsubsection{Uncertainty quantification}
Ill-posed inverse problems inherently possess multiple valid solutions for a single sparse observation. Our generative framework naturally accommodates this by providing principled Uncertainty Quantification (UQ). Because the guided integration is a deterministic mapping conditioned on a stochastic initial state, we can sample an ensemble of $K$ independent initial noise vectors $\{\mathbf{u}_0^{(k)}\}_{k=1}^K \sim \mathcal{N}(\mathbf{0}, \mathbf{I})$. Solving the guided ODE for each initialization yields a distribution of reconstructed physical fields $\{\mathbf{u}_1^{(k)}\}_{k=1}^K$. 
From this ensemble, we derive the pixel-wise empirical mean $\bar{\mathbf{u}}_1$ as the averaged ensemble prediction, and the standard deviation $\sigma_{\mathbf{u}_1}$ as a spatial map of epistemic uncertainty:
\begin{equation}
\bar{\mathbf{u}}_1 = \frac{1}{K} \sum_{k=1}^K \mathbf{u}_1^{(k)}, \quad \sigma_{\mathbf{u}_1} = \sqrt{\frac{1}{K} \sum_{k=1}^K \left( \mathbf{u}_1^{(k)} - \bar{\mathbf{u}}_1 \right)^2}
\end{equation}
This uncertainty map is highly interpretable, typically exhibiting lower variance near the observed boundaries and higher variance in uncertain internal regions, thus providing critical reliability bounds for downstream physical analysis.

\subsubsection{Physics decoder}
While the generated velocity fields describe the kinematic state of the granular flow, comprehensive analysis requires access to the underlying mechanics and kinetic measurements of velocity fluctuations, specifically the stress tensor and granular temperature. In granular physics, these macroscopic properties emerge from complex micro-scale interactions, including inter-particle friction, collisions, and force chains. The relationship between the mean velocity field and the stress state (i.e., the constitutive law) is often non-local and highly non-linear, making it difficult to derive analytically without assumptions for simplification or bounding.

To bridge the gap between the generated kinematics and granular mechanics without reverting to expensive DEM computations, we train a dedicated velocity-to-physics decoder, denoted as $\mathcal{H}_\psi$. \textcolor{black}{This model acts as a statistical regression model for diagnostic approximation rather than a universal constitutive surrogate. Because the macroscopic velocity field and its spatial gradients exhibit a strong statistical correlation with the local mechanical states in the studied inclined flows, the decoder effectively maps} the instantaneous velocity field directly to the stress state (mean normal stress $p$ and deviatoric stress $q$, as representatives to full stress tensor $\bm{\sigma}$), derived from particle contacts and momentum transfer, as well as the granular temperature $T$ (a measure of velocity fluctuation energy).

\textcolor{black}{We utilize a two-stage architecture rather than an end-to-end generative mapping from boundary velocity to internal stress. Physically, this preserves the relationship between kinematics and mechanics. Algorithmically, generating the intermediate velocity field allows the forward operator to map internal velocities to boundary velocities within the same physical domain, which facilitates stable gradient guidance. Direct generative modeling of highly stochastic stress fields from cross-domain boundary velocity observations would introduce severe optimization complexities.}

The physics decoder is defined as a deterministic mapping utilizing the same U-Net architecture as the forward operator elaborated in Section \ref{sec:forward_operator}. 
For a given velocity field slice $\mathbf{u}_{in}$ at a boundary-normal index $c$, the model predicts the co-located physical fields $p, q,$ and $T$:
\begin{equation}
[p, q, T] = \mathcal{H}_\psi(\mathbf{u}_{in}, c)
\end{equation}
The layer index $c \in \{0, 1, 2, 3\}$, which encompasses both interior and boundary slices, is embedded into the network. This conditioning is crucial because granular flows exhibit depth-dependent regimes, ranging from boundary-dominated slipping zones to bulk inertial flows, where the relationship between velocity gradients and stress varies significantly.

The decoder is trained in a supervised manner using paired samples from the dataset. We minimize the reconstruction error of the physical quantities: 
\begin{equation} 
\mathcal{L}_{phys}(\psi)=\mathbb{E}_{(\mathbf{u}, p, q, T, c) \sim \mathcal{D}} \left[ \left\| \mathcal{H}_\psi(\mathbf{u}, c) - [p,q,T] \right\|^2_2 \right]
\end{equation} Once trained, $\mathcal{H}_\psi$ allows us to instantaneously recover the stress state and fluctuation energy from the synthesized velocity fields, enabling a complete physical characterization of the granular system.

\subsection{Training strategy and performance measurement}

\subsubsection{Data preprocessing}
The training and evaluation data are sourced from 3D DEM simulations of gravity-driven granular flow on inclined planes. To ensure the model captures the diverse microstructural states, the dataset comprises six independent simulation instances generated with varying DEM random seeds for particle initialization. For the five runs designated for model development, 90\% of the data within each instance is randomly allocated for training, while the remaining 10\% is used for validation. The sixth simulation, initialized with a distinct random seed, is reserved for independent testing. To formulate the reconstruction task, the 3D volumetric data is discretized into a series of 2D boundary-parallel slices along the depth direction, i.e., the Slices 0-3 in Figure~\ref{fig:framework}(b), with Slice~0 as the observational boundary and Slices~1-3 as interior layers for reconstruction.

To ensure stable optimization of the neural networks, the raw kinematic and mechanical fields, including the velocity components ($v_x, v_y, v_z$), granular temperature ($T$), and stress invariants ($p, q$), are standardized using a robust global Z-score normalization. For any physical variable $\mathbf{x}$, the normalized input $\mathbf{x}'$ is computed as:
$
\mathbf{x}' = \frac{\mathbf{x} - \bm{\mu}_{\mathbf{x}}}{\bm{\sigma}_{\mathbf{x}}},
$
where the mean $\bm{\mu}_{\mathbf{x}}$ and standard deviation $\bm{\sigma}_{\mathbf{x}}$ are aggregated over the training dataset. Crucially, in granular systems, global averaging across the entire spatial domain would heavily skew the statistical moments toward zero due to the extensive presence of stationary particles, non-material spaces, and quasi-static zones. 
Thus, $\bm{\mu}_{\mathbf{x}}$ and $\bm{\sigma}_{\mathbf{x}}$ are computed exclusively over the active flow regime where the local velocity magnitude $|\bm{v}_{cell}| > 0.01$~m/s. 
These statistics are aggregated solely across the designated training configurations to prevent data leakage. During inference, the generated velocity field are denormalized back to their exact physical units for physical evaluation.

\subsubsection{Neural network configuration and training}\label{sec:training}
The generative flow matching backbone $f_\theta$ employs a U-Net architecture with channel dimensions of $[128, 256, 512, 512]$. Four-head attention mechanisms are integrated exclusively at the 512-channel layers to efficiently capture long-range spatial correlations. Virtual time $\tau$ and slice index $c$ are mapped via sinusoidal and learnable embeddings, respectively, and injected into all residual blocks through feature-wise addition.
The backbone is trained for \textcolor{black}{up to} 1000 epochs using AdamW. \textcolor{black}{To prevent overfitting and the associated risks of variance collapse or selective memorization of the training data \cite{dasgupta2026solving}, the vector field matching loss was monitored on the independent validation set. An early stopping criterion was employed to halt training when the validation loss plateaued, ensuring the generative prior preserves its diversity and generalizability.} The learning rate follows a cosine annealing schedule, decaying from a GPU-scaled base of $1 \times 10^{-4}$ to $5 \times 10^{-6}$. Gradients are clipped to a maximum norm of 1.0. 

The forward operator $g_\phi$ and physics decoder $\mathcal{H}_\psi$ share the same architecture, a deterministic conditional U-Net with channel dimensions of $[64, 128, 256, 512]$ and no attention modules. The slice index $c$ is embedded and injected into all residual blocks. Both networks are optimized via MSE loss for 500 epochs using AdamW. The learning rate follows a cosine annealing schedule, decaying from a GPU-scaled base of $2 \times 10^{-4}$ to $1 \times 10^{-6}$.
All models are implemented in PyTorch and trained using Distributed Data Parallel (DDP) on 4 H100 GPUs. 

During inference, the probability flow ODE is solved via the first-order Euler method with 100 steps ($\Delta \tau = 0.01$) and a guidance scale of $\xi = 1.0$. The consistency loss is evaluated within the observation window by passing the terminal state $\hat{\mathbf{u}}_1$ through the frozen forward operator $g$. Leveraging the spatially selective SSE formulation, the raw backpropagated gradient is applied directly without vector normalization, preserving the absolute physical scale of the measurement error and inherently preventing numerical divergence in zero-velocity space.

\subsubsection{Evaluation metrics}
The reconstructed fields are evaluated against the DEM ground truth and benchmarked against a deterministic CNN baseline. The evaluation utilizes either the ensemble mean of multiple generative trajectories for macroscopic structural assessment, or individual generative realizations to preserve fine-scale kinematic textures.
To prevent error dilution from non-material quasi-static space, statistical metrics are computed exclusively over active flowing regions. 
The absolute pixel-wise reconstruction error is quantified using the Root Mean Square Error (RMSE):
\begin{equation}
RMSE = \sqrt{\frac{1}{N_{act}}\sum_{i} M_i (y_i - \hat{y}_i)^2}
\end{equation}
where $y_i$ and $\hat{y}_i$ are the ground truth and predicted values at spatial index $i$, respectively. The term $M_i \in \{0, 1\}$ is a binary indicator mask that equals $1$ if the pixel belongs to the active flowing region (e.g., $|\bm{v}_{cell}| > 0.01$ m/s) and $0$ otherwise. Consequently, $N_{act} = \sum_i M_i$ represents the total number of active pixels within the evaluated domain.

Furthermore, to evaluate the spatial distribution alignment independent of absolute magnitudes, we compute the spatial Pearson correlation coefficient ($r$):
\begin{equation}
r = \frac{\sum_{i} M_i (y_i - \bar{y}_{act})(\hat{y}_i - \bar{\hat{y}}_{act})}{\sqrt{\sum_{i} M_i (y_i - \bar{y}_{act})^2} \sqrt{\sum_{i} M_i (\hat{y}_i - \bar{\hat{y}}_{act})^2}}
\end{equation}
where $\bar{y}_{act}$ and $\bar{\hat{y}}_{act}$ are the mean values of the ground truth and prediction within the masked region.
In the subsequent analyses, this correlation is evaluated independently for each spatial velocity component ($v_x, v_y, v_z$), with the respective coefficients denoted as $r_x, r_y,$ and $r_z$.

\section{Results}
The fundamental premise of the proposed framework relies on the backbone's ability to act as a robust kinematic prior, capable of synthesizing physically valid velocity fields independent of specific boundary constraints. 
To assess this unconditional generation capability, the probability flow ODE is integrated directly from standard Gaussian noise without any gradient guidance ($\xi = 0$). The resulting fields are statistically compared against the 10\% validation dataset derived from the DEM simulations.

\begin{figure}[H]
\centering
\includegraphics[width=\textwidth]{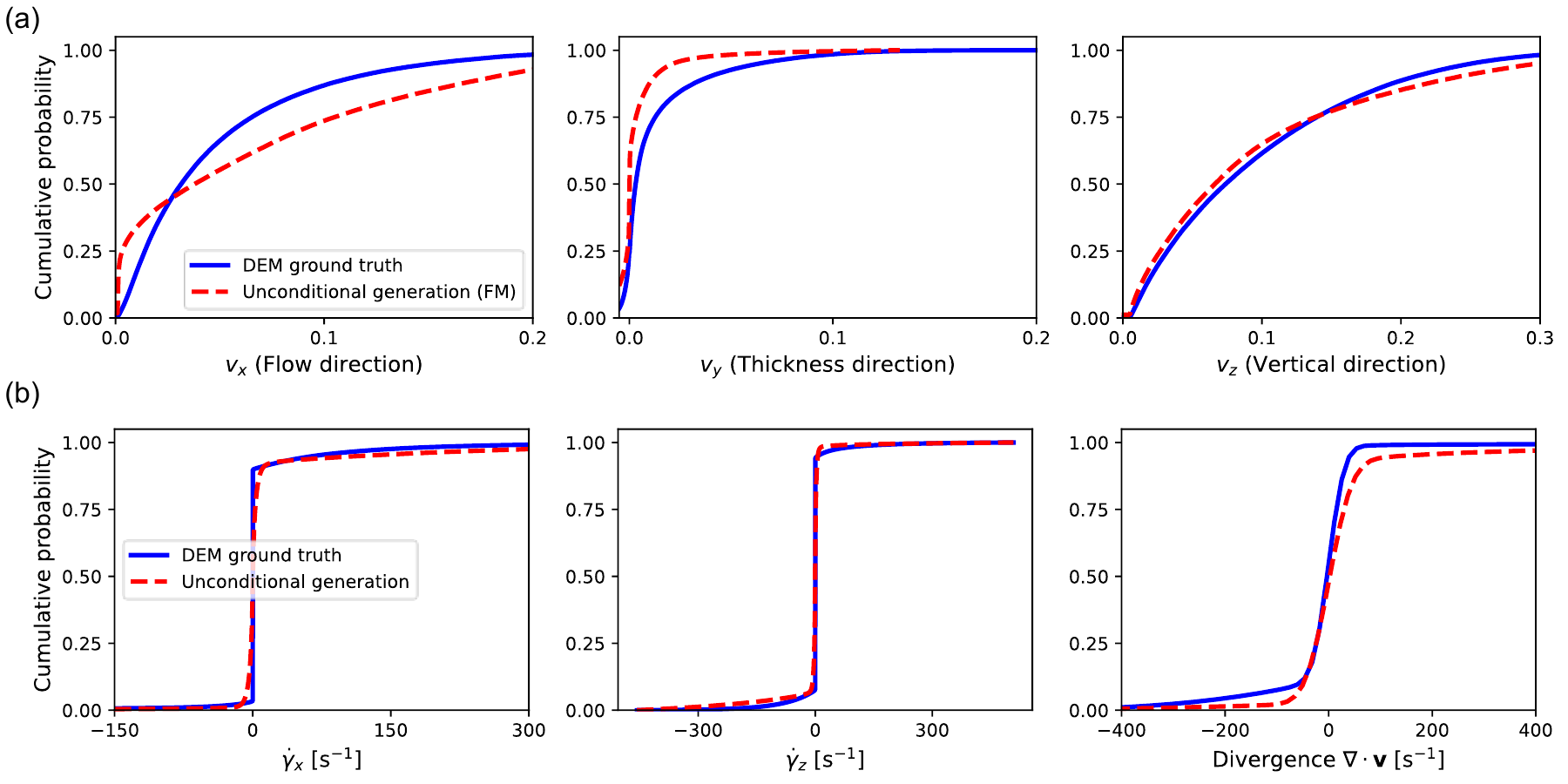}
\caption{\textcolor{black}{Validation of the generative kinematic prior. (a)} Empirical cumulative distribution functions (eCDFs) of the three velocity components ($v_x$, $v_y$, and $v_z$). \textcolor{black}{(b) Cumulative probability distributions of spatial velocity gradients, including shear strain rates ($\dot{\gamma}_x, \dot{\gamma}_z$) and volumetric divergence ($\nabla \cdot \mathbf{v}$).} The unguided generative samples show a strong overall match with the DEM ground truth, demonstrating that the model successfully captures \textcolor{black}{both} the underlying kinematic distribution \textcolor{black}{and spatial-physical constraints.}}
\label{fig:unconditional_pdf}
\end{figure}

Figure \textcolor{black}{\ref{fig:unconditional_pdf}(a)} illustrates the empirical cumulative distribution functions (eCDFs) of all three velocity components ($v_x, v_y, v_z$) for both the unconditionally generated ensemble and the numerical simulation dataset. 
Across all three directions, the primary bulk regions of the generative distributions exhibit good alignment with the DEM ground truth. 
However, noticeable deviations exist at the extreme upper and lower percentiles. This discrepancy is attributed to the unbounded Gaussian noise used in the generative modeling, which produced soft statistical "long tails" in the probability distribution, whereas the DEM data is governed by strict physical cutoffs (e.g., the hard zero bound and maximum velocities limited by gravity and friction). 

\textcolor{black}{While the marginal eCDF confirms the accuracy of the pixel-wise numerical ranges, physical validity also requires preserving spatial continuity. To verify this, Figure \ref{fig:unconditional_pdf}(b) evaluates the cumulative distributions of spatial velocity gradients, specifically the kinematic shear strain rates ($\dot{\gamma}_x, \dot{\gamma}_z$) and the volumetric divergence ($\nabla \cdot \mathbf{v}$). Because spatial gradients depend on adjacent pixels, they effectively measure local multi-point correlations. The gradient distributions of the generated fields closely match the DEM ground truth, successfully capturing the sharp transitions near zero that characterize quasi-static granular behavior. This confirms that the generative prior preserves essential spatial structures rather than merely generating independent pixels.}

Overall, the statistical \textcolor{black}{and spatial} congruence across the primary flow regimes is significant. This confirms that the FM backbone is not merely interpolating training data, but has effectively parameterized the underlying kinematic distribution of the bulk flow, establishing a robust physical prior for the subsequent observation-guided inference.

\subsection{Internal velocity fields reconstruction from full boundary observation}\label{sec:results_full_obs}
We first evaluate reconstruction of unobservable internal velocity fields from full boundary measurements using the proposed CFM framework coupled with the forward operator. The reconstruction is conditioned on the streamwise ($v_x$) and vertical ($v_z$) velocity profiles observed across the front boundary (Slice 0). Note that the wall-normal velocity component ($v_y$) is naturally unobservable at this boundary due to the physical wall constraints.

\begin{figure}[H]
\centering
\includegraphics[width=\textwidth]{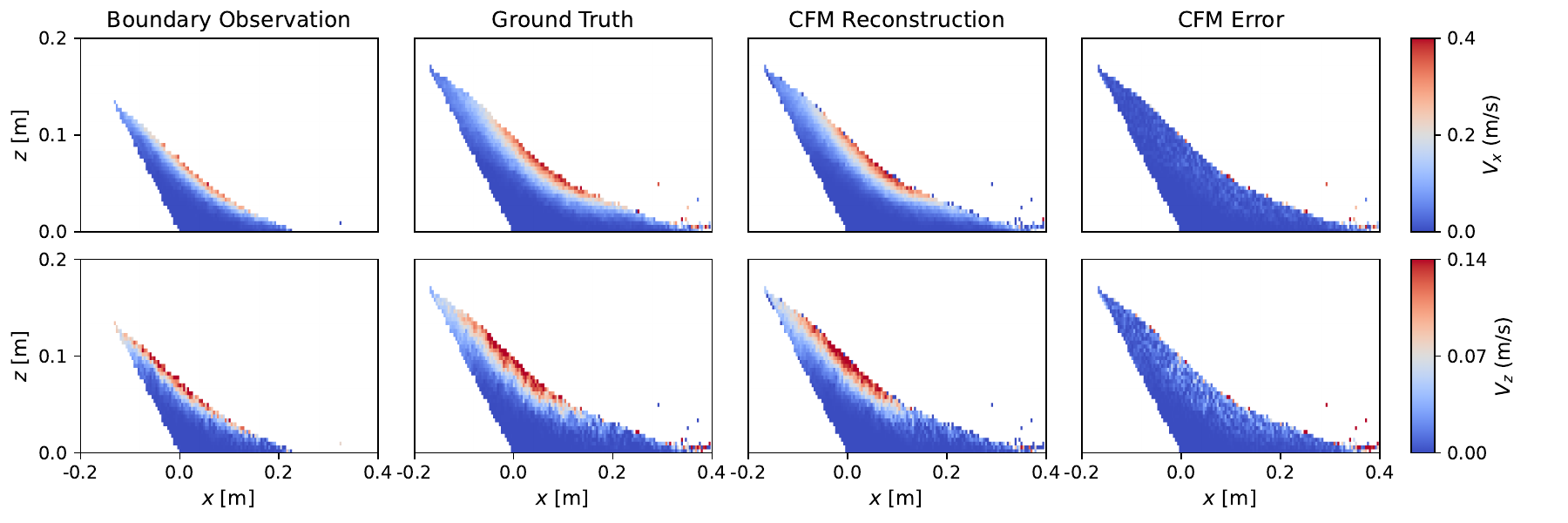}
\caption{Reconstruction of internal velocity fields from full boundary observations at $t = 0.56$~s. The top and bottom rows display the streamwise ($v_x$) and vertical ($v_z$) velocity components, respectively. From left to right, the columns present the boundary observation (Slice 0) used as the conditioning input, the DEM ground truth at the deepest internal plane (Slice 3), the corresponding CFM reconstruction, and the absolute error field.}
\label{fig:recon_velocity_full}
\end{figure}

While the CFM framework reconstructs the three-direction internal velocities at multiple slices for all time frames, Figure \ref{fig:recon_velocity_full} presents a representative snapshot at $t = 0.56$~s to illustrate the reconstruction performance. For clarity, only the observable flow components ($v_x$ and $v_z$) are shown at the deepest internal plane (Slice 3), which is the most difficult case because it is farthest from the observation boundary.

As demonstrated in Figure~\ref{fig:recon_velocity_full}, the reconstructed fields agree well with the DEM reference. The model captures the main flow structures, including the high-velocity region near the front and the depth-dependent shear pattern. The agreement remains good even at Slice 3, indicating that the learned prior and gradient guidance can propagate information from the observed boundary into the bulk while preserving the overall flow profile.

This visual agreement is supported by the quantitative metrics in the active region ($|\bm{v}_{cell}| > 0.01$\,m/s). At $t = 0.56$~s, the reconstructed $v_x$ and $v_z$ fields reach spatial correlations of $r_x = 0.878$ and $r_z = 0.729$, with RMSE values of 0.049 and 0.034~m/s, respectively. These results indicate that the framework recovers both the dominant spatial patterns and the velocity magnitudes with good accuracy.
Furthermore, to demonstrate the temporal robustness of the framework, reconstructions at additional flow stages ($t = 0.16$\,s and $t = 0.36$\,s) are provided in Appendix Figure~\ref{fig:appendix1}, consistently showing high fidelity across the entire flow duration.

Ultimately, while the generative framework demonstrates high fidelity when provided with dense and continuous boundary data, such ideal sensor coverage is rarely achievable in practical engineering scenarios. The true challenge lies in highly constrained sensing environments where measurements are sparse or incomplete. Therefore, the structural robustness of the proposed model is further evaluated under the severely ill-posed condition of partial observations in the subsequent section.

\subsection{Internal velocity fields reconstruction under partial boundary observations}\label{sec:results_partial_obs}
\begin{figure}[h!]
\centering
\includegraphics[width=\textwidth]{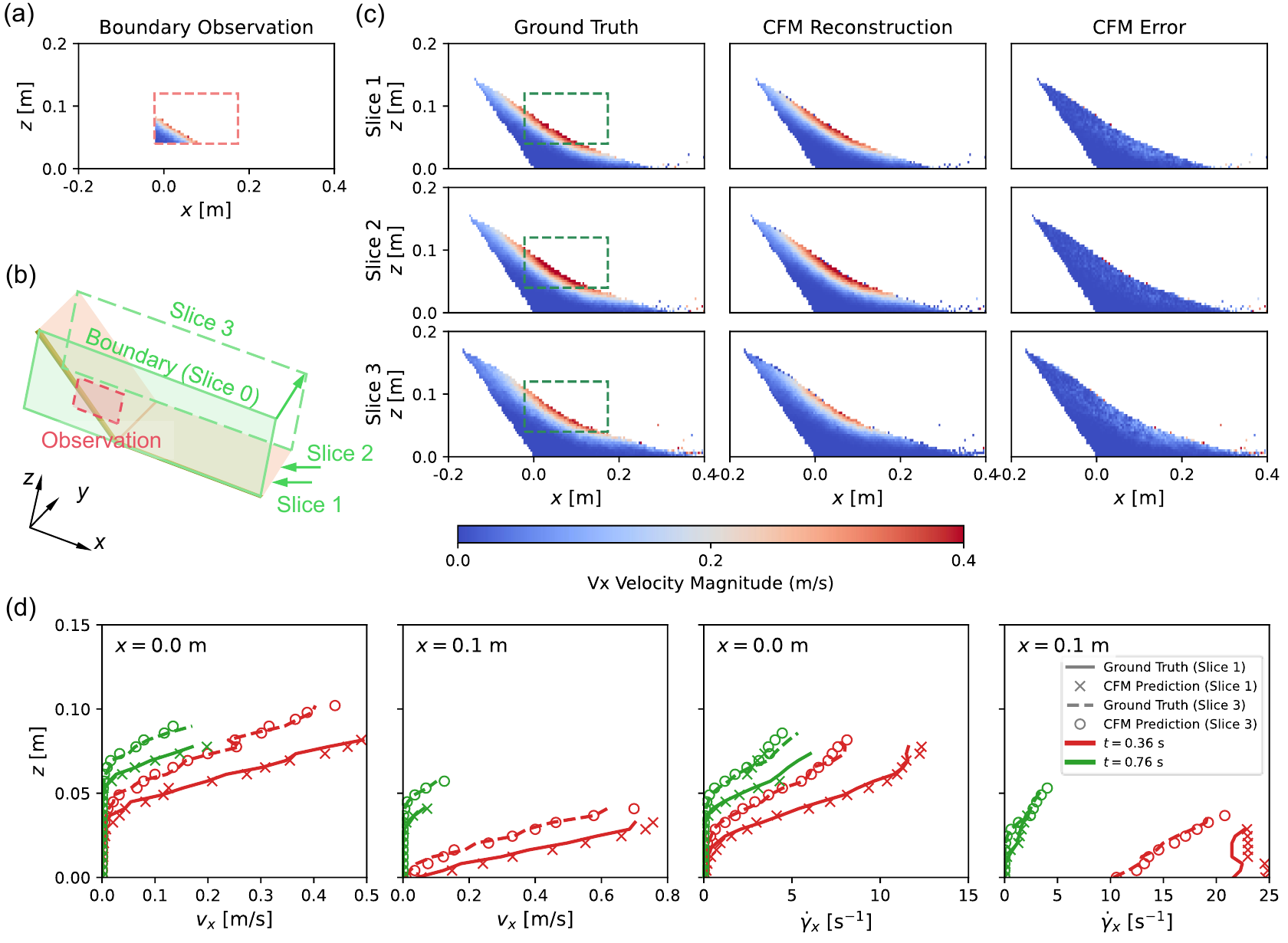}
\caption{Reconstruction of internal streamwise velocity fields ($v_x$) from partial boundary observations at $t = 0.56$~s. (a) The masked input boundary observation region, indicated by a pink bounding box. (b) Visualization of the observation window on the boundary (Slice 0) and prediction Slices 1-3. (c) Slice-by-slice comparison across three internal depths (rows: Slices 1, 2, and 3 from top to bottom). The columns display the DEM ground truth with green boxes showing the projection of the boundary observation window, the CFM reconstruction, and the absolute error, respectively. \textcolor{black}{(d) Vertical profiles of velocity ($v_x$) and shear strain rate ($\dot{\gamma}_x$) at $x = 0.0$~m and $x = 0.1$~m across two flow stages ($t = 0.36$ and $0.76$~s) along $z$ direction. The plots compare the ground truth and the CFM prediction at two different internal depths (Slice 1 and Slice 3).}}
\label{fig:recon_velocity_partial}
\end{figure}

To evaluate the framework under severely ill-posed conditions, the boundary observation (Slice 0) is restricted to a central patch, as highlighted by the pink box in Figure \ref{fig:recon_velocity_partial}(a). The corresponding 3D view in Figure~\ref{fig:recon_velocity_partial}(b) provides the geometric context of the flow, observation, and slice locations. Under this setting, only 17.4\% of the globally active particles are covered by the observation window, and the head and tail regions are completely unobserved.

Figure \ref{fig:recon_velocity_partial}(c) presents the reconstructed $v_x$ fields at $t = 0.56$~s for Slices 1–3. Inference for the $v_y$ and $v_z$ is omitted for brevity. The green boxes in the DEM panels indicate the spatial projection of the observed boundary window onto each interior slice. The material distribution and velocity magnitude in the green boxes of Slices 1-3 are different from each other and from the observation (Slice 0), indicating that the observed region on the boundary does not directly reveal the full structure of the interior flow, especially as depth increases. 

Despite the limited input, the generative model successfully recovers the material distribution of the interior flow across all three slices. In particular, it reconstructs the unobserved front and tail regions and maintains coherent depth-wise evolution from Slice 1 to Slice 3. As expected, the reconstruction quality decreases with distance from the observed boundary, but the deterioration is marginal.

The quantitative results confirm this trend. For $v_x$, the spatial correlations are 0.953, 0.932, and 0.868 for Slices 1, 2, and 3, respectively, with corresponding RMSE values of 0.038, 0.044, and 0.055~m/s. These results indicate that the method remains effective even when the observation is sparse and spatially localized.

The error maps (on the right of Figure~\ref{fig:recon_velocity_partial}(c)) also show local discrepancies near the flow front, where particle splashing and intermittent motion are most pronounced. Such regions are intrinsically difficult to match pointwise because their dynamics are highly sensitive to local particle arrangements with stochasticity. Even so, the reconstructions remain physically plausible and preserve the large-scale kinematic structure.

\textcolor{black}{To further examine the depth-wise kinematic structure, Figure~\ref{fig:recon_velocity_partial}(d) presents the $z$-direction profiles of velocity ($v_x$) and shear strain rate ($\dot{\gamma}_x$) at two representative locations ($x = 0.0$~m and $x = 0.1$~m) across two flow stages ($t = 0.36$ and $0.76$~s). 
Although the flow exhibits significant spatial and temporal variations from the shallow region (Slice 1) to the deep interior (Slice 3), the CFM predictions consistently align well with the internal ground truth. 
The model accurately reproduces the bulk flow velocity across different depths; while it successfully captures the nonlinear local velocity gradients at the later stage ($t=0.76$~s), some deviations are observed in the shear rate predictions during the earlier transient phase ($t=0.36$~s).
}

Overall, these results show that the proposed framework can infer interior flow fields under severely incomplete observations, which is the regime of greatest practical interest.

\subsection{Inference of physical fields}
Besides the kinematic velocity fields, internal stress fields (stress tensor $\bm{\sigma}$ or representative stress invariants $p$ and $q$) and localized velocity fluctuation (granular temperature $T$) are also critical in order to understand the flow behavior. These fields reflect the mechanical states and energy dissipation characteristics of the material and are directly connected to granular rheology.

To demonstrate the capability to predict these physical fields, the pre-trained physics operator is utilized to map the conditionally generated velocity fields directly to these physical quantities. This represents an end-to-end inference task: inferring deep internal physics solely from the severely masked boundary observations (the same 17.4\% window setup as in Section~\ref{sec:results_partial_obs}).

\begin{figure}[H]
\centering
\includegraphics[width=\textwidth]{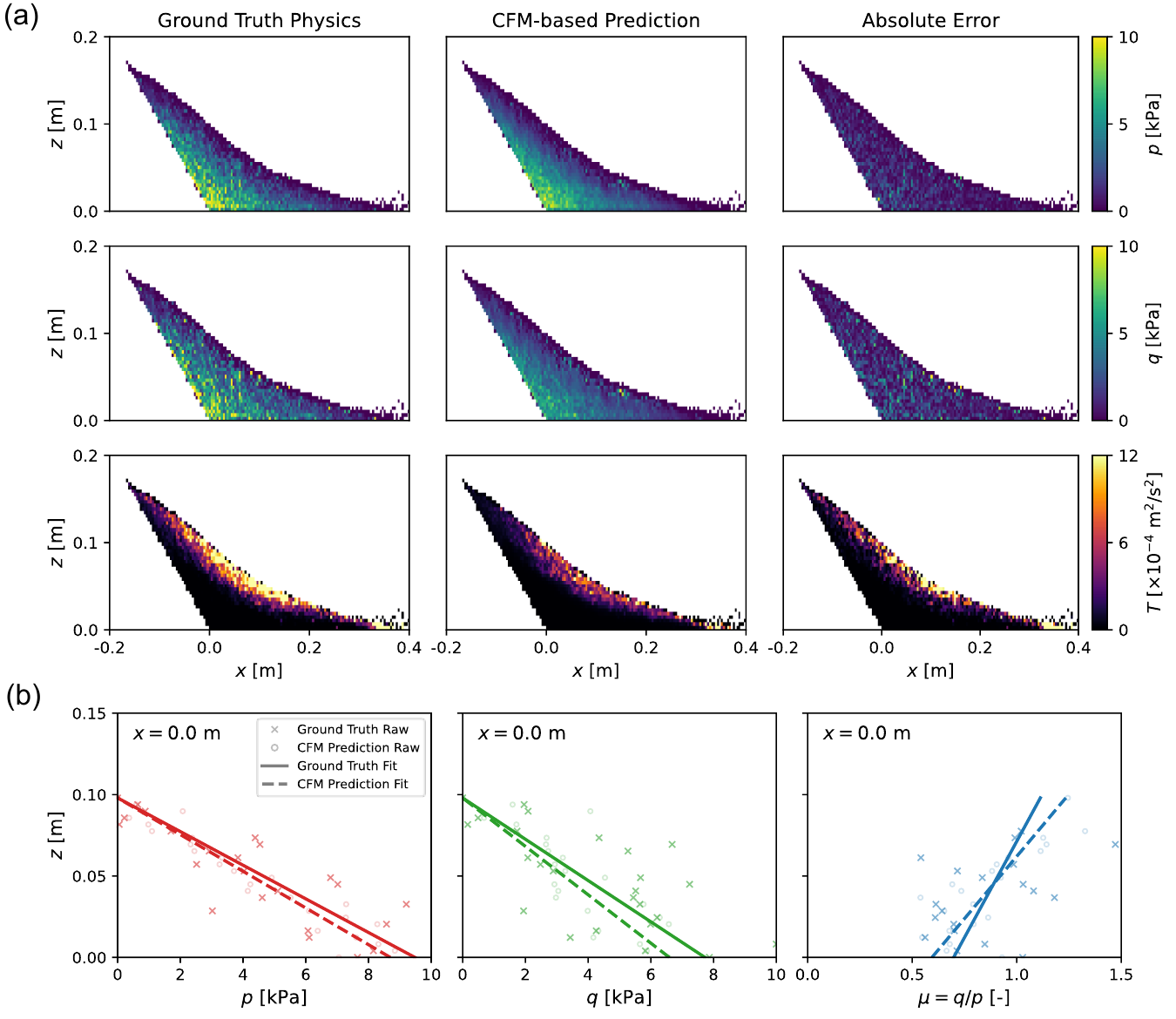}
\caption{End-to-end inference of physical fields from partial boundary observations at $t = 0.56$ s \textcolor{black}{for Slice 3}. \textcolor{black}{(a) Spatial distributions of} pressure $p$, shear stress $q$, and granular temperature $T$ at the deepest unobservable plane. The columns display the DEM ground truth, the chained CFM prediction, and the absolute spatial error, respectively. \textcolor{black}{(b) Depth-wise profiles of $p$, $q$, and the effective friction coefficient ($\mu = q/p$) at $x = 0.0$~m, comparing the raw data and linear fits between the ground truth and CFM predictions.}}
\label{fig:physics_inference}
\end{figure}

Figure \ref{fig:physics_inference}(a) visualizes the spatial distributions of $p$, $q$, and $T$ for the deepest unobservable plane at $t = 0.56$~s. To systematically evaluate the inference quality, the visualization is structured as a 3$\times$3 grid. The rows correspond to the three physical quantities ($p$, $q$, and $T$) from top to bottom. The columns systematically compare the DEM ground truth, the chained CFM prediction, and the absolute spatial error. 
Visual inspection confirms a strong structural agreement between the variable fields inferred from the CFM-generated velocity and the ground truth. The model effectively captures the dense compressive regions (high pressure $p$), the localized energy transfer and shearing pathways (deviatoric stress $q$), and the active kinetic layers explicitly represented by the granular temperature $T$.

The statistical metrics reveal a consistent performance. The reconstructed macroscopic pressure field ($p$) achieves a spatial correlation of $r = 0.832$ and an RMSE of 1.43~kPa. The deviatoric stress ($q$) is inherently tied to higher-order spatial gradients and thus sensitive to pixel-wise mismatches, naturally exhibiting a moderately higher RMSE of 2.12~kPa but successfully preserving the global distribution with $r = 0.600$.

Notably, for the granular temperature ($T$), a quantity directly relying on stochastic velocity fluctuations, the model achieves a remarkably low RMSE of $1.22\times10^{-3}$ m$^2/$s$^2$. While the framework successfully tracks the macroscopic kinetic energy, the spatial correlation mathematically drops to $r = 0.322$ for this specific time frame. This is attributed to a well-known statistical artifact: as the granular avalanche decelerates and approaches a near-static state in certain unobserved regions, the true physical fluctuations diminish toward zero. Consequently, the Pearson correlation becomes highly sensitive to infinitesimal numerical noise, driving down the $r$ value despite the outstanding absolute accuracy characterized by the low RMSE.
Overall, these metrics demonstrate that the generative framework effectively infers the mechanics and thermodynamics of granular media directly from blind boundaries.

\textcolor{black}{To complement the spatial maps, Figure~\ref{fig:physics_inference}(b) examines the depth-wise profiles of $p$, $q$, and the effective friction coefficient ($\mu = q/p$) at $x = 0.0$~m for the same time frame ($t = 0.56$~s). To account for the inherent noise in the raw granular data, linear fits are applied to extract the macroscopic trends. The CFM fitted lines accurately capture the overall trend of the linear increase in pressure along the depth and the corresponding deviatoric stress variations. Nevertheless, localized deviations persist as an inherent limitation of data-driven generative models.}

\textcolor{black}{To isolate the sources of error in the mechanical reconstruction, a separate evaluation was performed by feeding the ground-truth DEM velocity fields directly into the physics decoder. This separates the decoder's inherent statistical approximation error from the errors propagated through velocity reconstruction. Results show that the inherent approximation of the decoder accounts for the vast majority of the final prediction error. At $t$ = 0.36 s, using ground-truth velocities yields an active RMSE of 1.91~kPa for pressure $p$ and 2.63~kPa for deviatoric stress $q$. When using CFM-reconstructed velocities, the total cascaded errors increase only marginally to 2.07~kPa and 2.87~kPa, respectively. This tight margin is equally pronounced at $t$ = 0.56~s (Figure \ref{fig:physics_inference}). This confirms that the physics decoder is highly capable of executing the kinematics-to-mechanics mapping without catastrophically amplifying upstream velocity inaccuracies.}

\section{Analyses and discussion}

\subsection{Comparison with deterministic model}
\label{sec:CNN}
\subsubsection{Baseline CNN Specifications}
To benchmark the developed generative framework, a deterministic Convolutional Neural Network (CNN) is employed as a baseline to directly learn the inverse mapping from the boundary observation to the internal velocity field. To ensure a strictly fair comparison, the baseline CNN adopts a conditional U-Net architecture identical in capacity to the deterministic Forward Operator used in our framework.

The baseline model takes a 2-channel input, corresponding to the observed $v_x$ and $v_z$ fields at the boundary, and predicts the 3-component velocity field ($v_x, v_y, v_z$) at the target internal depths. The encoder-decoder structure utilizes four scale levels with feature map dimensions of $[64, 128, 256, 512]$. Each level consists of residual blocks equipped with Group Normalization (8 groups) and SiLU activation functions. Spatial downsampling is achieved via strided convolutions, while upsampling utilizes transposed convolutions with skip connections concatenating symmetric encoder features.
Additionally, because the inverse mapping is depth-dependent, the target internal slice index $c$ is treated as a conditional input. This depth indicator is passed through a multi-layer perceptron (MLP) with GELU activations to generate a dense class embedding, which is then spatially broadcast and added to the intermediate feature maps within every residual block.

The baseline model is trained on the same normalized dataset and train-validation splits as the CFM framework, using pixel-wise MSE loss and AdamW optimizer with a cosine annealing learning rate scheduler (decaying to $10^{-6}$) for 1000 epochs.

\subsubsection{Comparison of reconstruction fidelity}
\begin{figure}[h!]
\centering
\includegraphics[width=\textwidth]{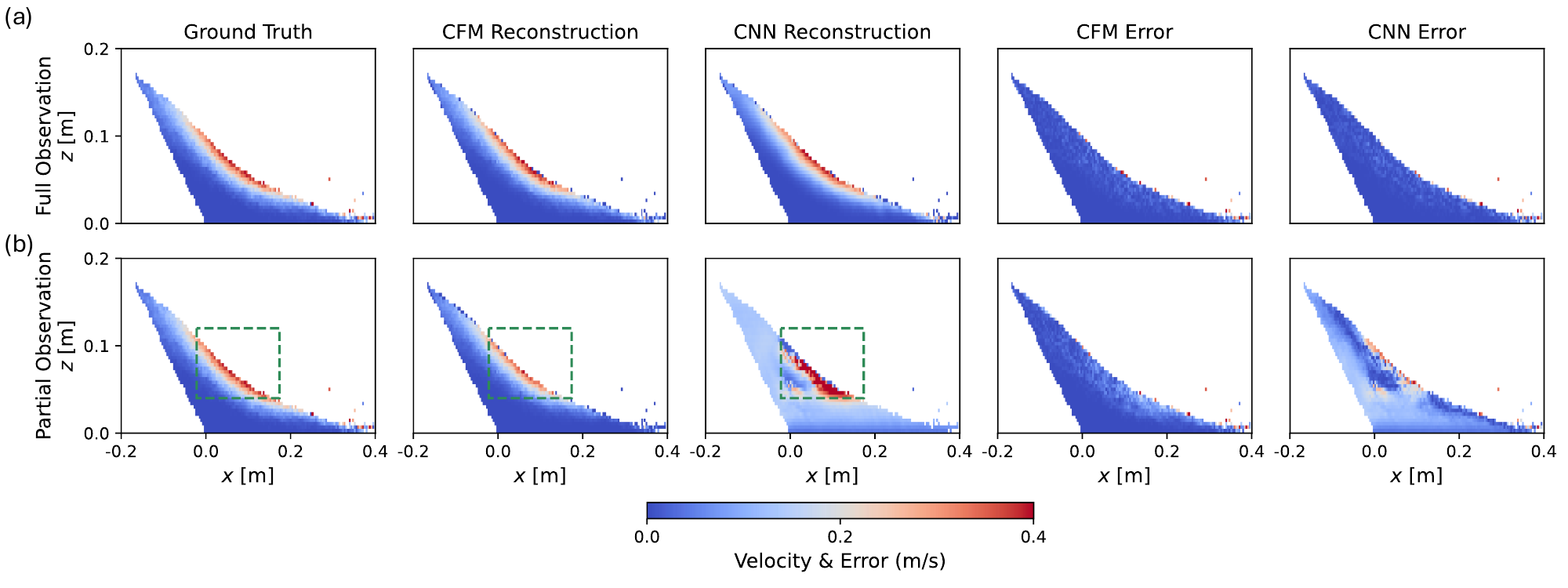}
\caption{Comparison of the reconstructed streamwise velocity ($v_x$) between the generative CFM framework and the deterministic CNN baseline at the deepest interior plane (Slice 3) for $t = 0.56$~s. The columns display the DEM ground truth, CFM and CNN predictions, and their respective absolute error maps. (a) Results under full boundary observation. (b) Results under severe partial observation.}
\label{fig:cnn_comparison}
\end{figure}

Figure \ref{fig:cnn_comparison} compares the proposed CFM framework with the deterministic CNN baseline for reconstruction of the streamwise velocity $v_x$ at the deepest interior slice (Slice 3). The top row corresponds to full boundary observation, and the bottom row to partial observation. The columns show the DEM reference, CFM reconstruction, CNN prediction, and corresponding absolute error maps.

Under full boundary observation (Figure~\ref{fig:cnn_comparison}(a)), both models perform well. The main flow structure is recovered in both cases, although the CNN prediction is slightly smoother. 
Quantitatively, the CFM framework achieves a spatial Pearson correlation of $r_x = 0.887$ and an RMSE of 0.087~m/s, compared with $r_x = 0.874$ and an RMSE of 0.091~m/s for the CNN. The difference is modest, which is expected when the observation is already highly informative.

The difference becomes much clearer under the severely ill-posed partial observation condition (Figure~\ref{fig:cnn_comparison}(b)). Green boxes denote the spatial regions on the inner slices corresponding to the boundary observation window (same as shown Figure~\ref{fig:recon_velocity_partial}). When the input is restricted to a localized central patch, the deterministic CNN experiences a significant drop in performance, with a spatial correlation decreasing to $r_x = 0.391$ and RMSE increasing to 0.174~m/s. In contrast, the CFM framework maintains a robust spatial correlation of $r_x = 0.875$ and a lower RMSE of 0.095~m/s. 
The visual comparison is consistent with these metrics: the CNN fails to reconstruct the unobserved head and tail regions and instead produces a blurred, unphysical mean state, while the CFM model recovers plausible kinematics throughout the hidden region. 

This contrast reflects the difference \textcolor{black}{in distributional behavior} between deterministic regression and conditional generative inference. A \textcolor{black}{deterministic} CNN trained with MSE tends to return a conditional average when the inverse problem is underdetermined, which leads to over-smoothing in regions with high uncertainty. \textcolor{black}{This averaging effect inherently over-smooths the flow fronts and misses localized high-fluctuation regions.} Conversely, \textcolor{black}{an individual sample from} the CFM model \textcolor{black}{acts as a direct draw from the learned conditional distribution}. It reconstructs \textcolor{black}{a specific,} mechanically valid velocity field instead of statistical averages, preserving \textcolor{black}{the localized kinetic textures, fine-scale variability,} and coherence in completely unobserved regions.

Overall, the comparison shows that deterministic regression can be competitive when the observation is nearly complete, but it becomes unreliable when information is sparse. The advantage of the generative formulation is most apparent in the ill-posed regime that motivates this study. Furthermore, this comparison also highlights the critical need to evaluate prediction confidence and leads to the uncertainty quantification in the following section.

\subsection{Uncertainty quantification}
While predictive uncertainty exists even under full boundary observations due to the inherent stochasticity of granular interactions (aleatoric), it is substantially magnified under the partial-observation setting (epistemic), where the inverse problem is mostly ill-posed. In this case, multiple interior flow states can be consistent with the same boundary measurement, so reconstruction quality should be assessed together with predictive uncertainty. 

\begin{figure}[h!]
\centering
\includegraphics[width=\textwidth]{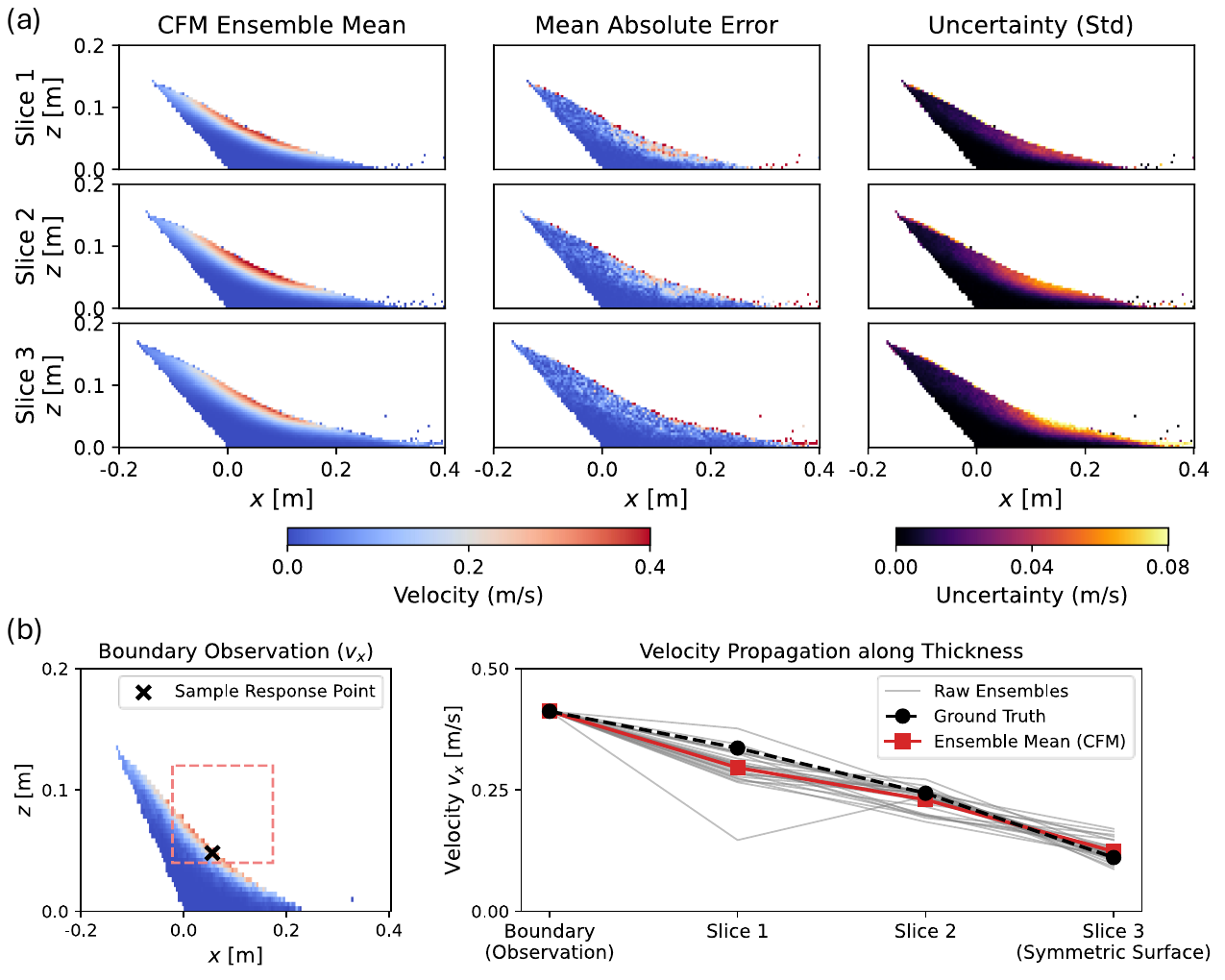}
\caption{Uncertainty analysis of the internal flow reconstruction under partial observation at $t = 0.56$~s. (a) Slice-by-slice spatial evaluation across three internal depths, displaying the CFM ensemble mean, absolute error, and pixel-wise standard deviation. (b) 1D velocity propagation profile along the thickness ($y$) direction extracted at a sample boundary location (marked by the cross `$\times$'). The plot compares the depth-wise ground truth velocity (black dashed line) against the CFM ensemble mean (red line) and the full distribution spread of all 25 raw generative ensembles (gray lines).}
\label{fig:recon_uq_partial}
\end{figure}

To this end, 25 independent inference passes (ensembles) are generated from different Gaussian initializations for the same partial observation at $t = 0.56$~s. Figure \ref{fig:recon_uq_partial}(a) shows the ensemble mean, its absolute error, and the predictive uncertainty (pixel-wise standard deviation) for $v_x$ in Slices 1 to 3. The DEM reference is omitted because it is the same as in Figure \ref{fig:recon_velocity_partial}.

The ensemble mean recovers the main flow profiles across all three slices. Quantitatively, the spatial Pearson correlations of the ensemble mean are $r_x = 0.963$, $0.957$, and $0.910$, with corresponding RMSEs of $0.036$, $0.038$, and $0.043$~m/s for Slices 1, 2, and 3, respectively. As expected, the accuracy decreases gradually with distance from the observation boundary (along the $y$ direction). This same trend is reflected in the uncertainty maps: the standard deviation increases with depth and is the largest near the unobserved tail region. \textcolor{black}{A direct visual comparison in Figure \ref{fig:recon_uq_partial}(a) confirms that the spatial distribution of predictive uncertainty generally correlates with the actual mean absolute error, although the uncertainty exhibits a physically expected conservative peak at the highly stochastic free surface.}

Interestingly, compared to the single realizations shown earlier in Section~\ref{sec:results_partial_obs}, the CFM ensemble mean appears smoother, naturally filtering out localized velocity fluctuations. This observation aligns with a fundamental statistical law: the expected value (mean) of a stochastic distribution inherently averages out spatial fluctuations. This directly explains why deterministic regression models (like the CNN baseline in Section~\ref{sec:CNN}, which minimizes MSE to find the conditional mean) inevitably suffer from over-smoothing. The CFM framework mitigates this limitation by allowing for both sharp single physical realizations and stable ensemble averages.

Figure \ref{fig:recon_uq_partial}(b) provides a local depth-wise velocity profile along the $y$ direction extracted at a representative boundary location. The ensemble mean follows the attenuation of velocity from the boundary to the deepest interior layer, and the spread of the 25 samples brackets the true value at each depth. At Slice 3, for example, the ground-truth velocity of 0.111~m/s lies within the predicted ensemble centered near 0.118~m/s. This local probe illustrates how uncertainty grows as the boundary constraint weakens with depth.

It may be initially counterintuitive that the velocity decreases from the observation boundary (Slice 0) to the other boundary (Slice 3) instead of showing a symmetric pattern. It is because the random particle initialization makes the inner domain (approaching Slice 3) have more particles (with a slightly higher initial packing) before the flow. This difference can be observed from Figure~\ref{fig:recon_velocity_partial}, where the green box in Slice 3 contains more material than Slices 2 and 1, with the Slice 0 contains the least. Therefore, the sampling point in \ref{fig:recon_uq_partial}(b), although on the top surface of Slice 0 with a high velocity, becomes under the top surfaces for Slices 1-3 (with the point-to-the-top distance rises as $y$ increases), resulting in the velocity magnitude decreasing (shown in Figure \ref{fig:recon_uq_partial}(b)). Consequently, this is a physical reconstruction aligning with the ground-truth DEM modeling.

The model robustness \textcolor{black}{and the quantitative calibration of the predicted uncertainty are} further evaluated through a statistical coverage analysis. Across the active flowing regions, 93.2\% of the ground-truth velocities fall within the $\pm 2\sigma$ confidence interval from the ensemble mean, and 97.9\% fall within the $\pm 3\sigma$ interval. These values indicate that the predicted spread is well aligned with the observed error distribution. It does not over-confidently generate incorrect structures, nor does it yield overly conservative and uninformative variances. This serves as a powerful demonstration of the generative prior: the model does not merely memorize the geometric location of the observation window. Instead, it actively acknowledges the physical possibility of complex unobserved dynamics, such as residual particle settling in the hidden tail domain, thereby providing reliable statistical boundaries for downstream engineering analysis. This capability is important for granular-flow inference, where incomplete measurements and non-unique interior responses are intrinsic to the problem.

\textcolor{black}{It is also worth noting that generative models trained on finite or narrowly distributed datasets can theoretically suffer from degeneracy, such as variance collapse or selective memorization of the training data \cite{dasgupta2026solving}. In such degenerate cases, the model either collapses to a deterministic mean or merely regurgitates specific training instances, losing its generative diversity. The well-calibrated statistical coverage demonstrated above confirms that the present CFM framework successfully avoids these weaknesses, maintaining physically meaningful variance even when trained on a restricted set of DEM configurations.}

\subsection{Ablation study}
\subsubsection{Effectiveness of the proposed gradient guidance}
To validate the proposed guidance, we compare it against \textcolor{black}{two} conventional baseline guidance \textcolor{black}{strategies}: \textcolor{black}{1) a standard normalized guidance} widely utilized in continuous fluid generation \cite{lipman2022flow, parikh2025conditional}, \textcolor{black}{and 2) a standard Mean Squared Error (MSE) formulation}.
Standard generative inference typically normalizes the guidance gradient to prevent numerical divergence, scaling it by the $L_2$ norm of the predicted vector field $f_\theta$:
\begin{equation}
\tilde{f}_\tau(\mathbf{u}_\tau, \tau, c) = f_\theta(\mathbf{u}_\tau, \tau, c) - \xi \left( \frac{\nabla_{\mathbf{u}_\tau} \mathcal{L}}{\| \nabla_{\mathbf{u}_\tau} \mathcal{L} \|^2_2 + \epsilon} \right) \| f_\theta(\mathbf{u}_\tau, \tau, c) \|^2_2
\label{eq:baseline_guidance}
\end{equation}
where $\epsilon$ is a small constant for numerical stability. 
In non-material empty regions, the normalization term ($\| \nabla_{\mathbf{u}_\tau} \mathcal{L} \|^2_2$) amplifies small numerical noise, generates large, unphysical velocity corrections in empty grids. \textcolor{black}{Alternatively, adopting a standard MSE formulation divides the total squared error by the total number of pixels. In granular flows dominated by empty space, this spatial averaging artificially dilutes the correction gradient to near zero, stripping it of its absolute physical magnitude.} Conversely, our unnormalized formulation allows the local error gradient to naturally vanish when empty-pixel predictions match zero-velocity observations.

\begin{figure}[h!]
\centering
\includegraphics[width=\textwidth]{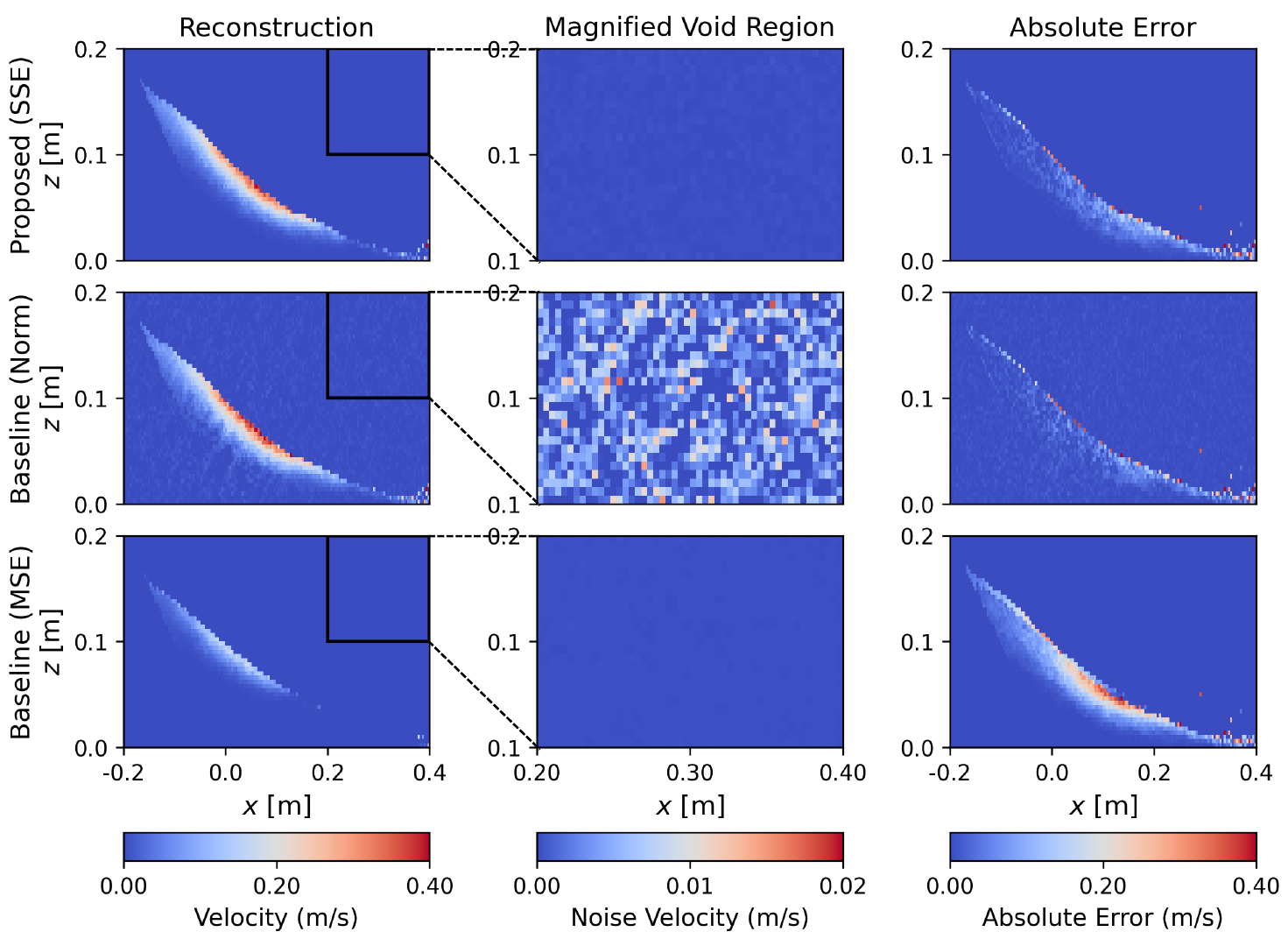}
\caption{Ablation study comparing the proposed sparsity-aware guidance (top row) \textcolor{black}{and two baselines: normalized guidance (middle row) and MSE guidance (bottom row)}. The columns display the reconstructed velocity field, a zoom-in view of the empty region, and the global absolute error map without background masking. The \textcolor{black}{normalized} baseline introduces severe non-physical numerical noise in empty and zero-velocity pixels and corrupts the active flow profile, \textcolor{black}{while the MSE baseline suffers from severe gradient dilution}. \textcolor{black}{In contrast}, the proposed method accurately maintains sharp material boundaries and absolute zero values in those spaces.}
\label{fig:ablation_guidance}
\end{figure}

Figure \ref{fig:ablation_guidance} compares the reconstructions obtained with the proposed sparsity-aware guidance (top row) and the baseline standard guidance (bottom row) at $t = 0.56$~s under the challenging partial observation condition. 
Note that the empty regions were visually masked in previous sections to highlight active flow, while the full domain is displayed here to show the difference and to evaluate the models' capability to delineate the sharp material and non-material interface, a ill-posed task for continuous neural networks, which inherently struggle with abrupt spatial discontinuities.

Results in Figure~\ref{fig:ablation_guidance} perfectly align with the above analysis. As shown in the zoom-in middle panels and the unmasked error maps in the right panel, the \textcolor{black}{normalized} baseline generates severe high-frequency disturbances and unphysical numerical predictions that propagate into the empty regions. \textcolor{black}{Meanwhile, the MSE baseline fails to dynamically adjust the vector field due to gradient dilution, resulting in massive absolute errors across the entire active domain.} This non-material, non-physical velocity field further compromises the fundamental mass conservation and exacerbates the mismatch with the discrete nature of granular materials. In contrast, the proposed guidance method effectively suppresses these unphysical fluctuations, maintaining sharp, well-defined boundaries between the active flowing mass and the surrounding empty space.

Quantitative metrics support this observation. Within the active flowing region, the proposed method achieves a spatial correlation of $r = 0.868$ and an RMSE of $0.100$~m/s. \textcolor{black}{The normalized baseline drops to $r = 0.106$ with an active RMSE of $0.258$~m/s, while the MSE baseline completely fails to recover the kinematic structure, yielding $r = 0.010$ and an active RMSE of $0.291$~m/s.} The difference is also clear in the non-material region: the proposed formulation gives an RMSE of $0.002$ m/s, compared with $0.010$ m/s for the \textcolor{black}{normalized} baseline and $0.007$ for the MSE baseline. When evaluating over the full domain, the global RMSE is $0.087$ m/s for the proposed method, \textcolor{black}{$0.254$}~m/s for the \textcolor{black}{normalized} baseline, \textcolor{black}{and $0.283$~m/s for the MSE baseline.}

These results demonstrate that preserving the absolute scale of the measurement discrepancy \textcolor{black}{via the unnormalized SSE formulation}, rather than applying artificial vector normalization \textcolor{black}{or spatial error averaging}, is critical for granular flows. It simultaneously enforces strict boundary consistency in the active flow zones and prevents unphysical predictions in the empty space.

\subsubsection{Robustness against restricted observation coverage and diluted resolution}\label{sec:ablation_obs}
This section details how reconstruction quality changes as the available boundary information is reduced. To isolate the effect of observation quality, this analysis \textcolor{black}{primarily} focuses on a representative challenging case: reconstruction of the deepest interior slice (Slice 3) at \textcolor{black}{$t = 0.16, 0.56,$ and $0.76$~s, which are representative for early-acceleration, relatively steady-flow, and deposition phases}. \textcolor{black}{Results for the shallower Slices 1 and 2 are provided in \ref{appendix2} to further demonstrate spatial representativeness.}

\begin{figure}[H]
\centering
\includegraphics[width=\textwidth]{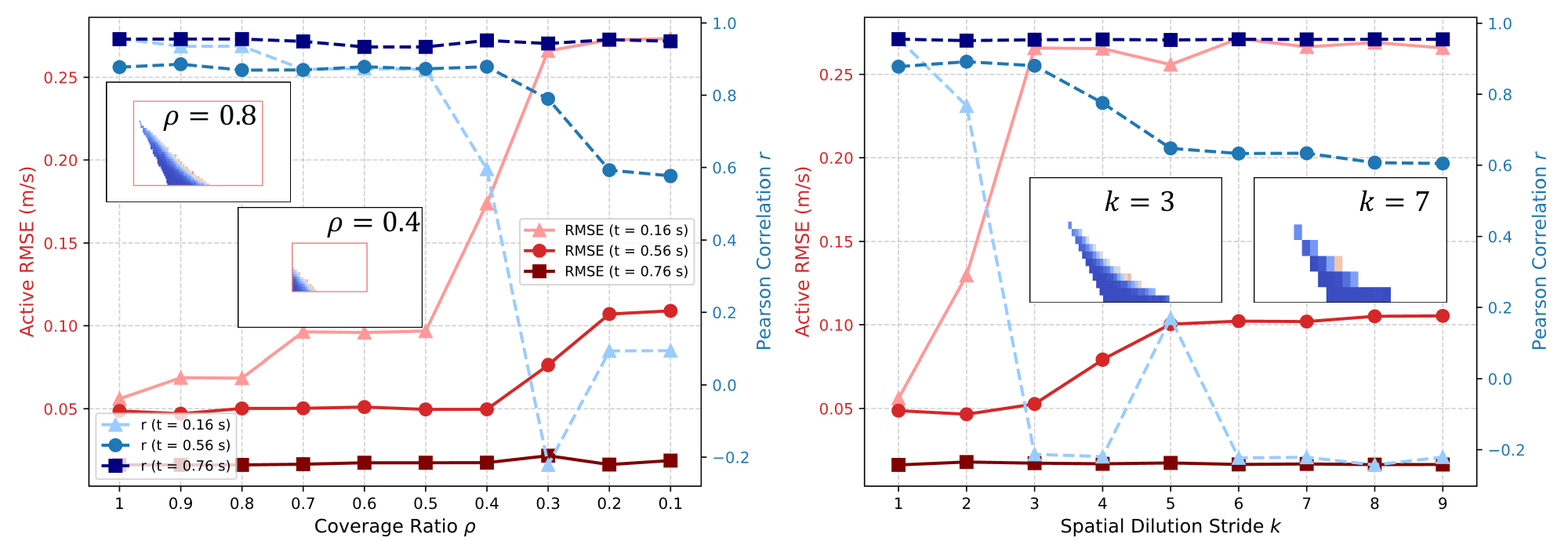}
\caption{Sensitivity analysis of the reconstruction fidelity under varying observation constraints (evaluated on Slice 3 at \textcolor{black}{$t = 0.16, 0.56,$ and $0.76$~s}). Left: Robustness against restricted spatial coverage, where the horizontal axis represents the varying coverage ratio $\rho$. Representative input conditions at $\rho = 0.8$ and $\rho = 0.4$ are visualized as insets. Right: Robustness against diluted spatial resolution, where the horizontal axis denotes the spatial subsampling stride $k$, with insets illustrating the sparse inputs at $k = 3$ and $k = 7$. In both plots, \textcolor{black}{solid lines with varying color shades} represent the active RMSE \textcolor{black}{at different flow stages}, and the \textcolor{black}{corresponding} dashed lines represent the spatial Pearson correlation $r$ (right vertical axis).}
\label{fig:ablation_combined}
\end{figure}

\textbf{Robustness against restricted spatial coverage.}
In industrial and experimental settings, the optical field of view may be obstructed, leading to severe spatial truncation of boundary measurements. This constraint is evaluated by varying the coverage ratio $\rho \in [0.1, 1.0]$. 
As shown in Figure \ref{fig:ablation_combined} (left), the CFM framework remains stable over a broad range of coverage loss. The spatial correlation $r$ maintains above 0.86 as $\rho$ decreases from 100\% to 40\%. Note that the window height and width are reduced simultaneously, so $\rho=0.4$ corresponds to only 16\% of the original observed area (as shown in the inset sub-figure). In this range, the RMSE changes only modestly, indicating that the main flow structure can still be recovered from a relatively small spatial anchor.

A clearer deterioration appears when the coverage is reduced below $\rho = 0.4$. \textcolor{black}{Taking $t=0.56$~s as an example,} at $\rho = 0.2$, the RMSE increases from $0.050$ to $0.107$~m/s, and the correlation also decreases significantly. \textcolor{black}{These sensitivity trends hold consistently across all evaluated flow stages ($t = 0.16, 0.56,$ and $0.76$~s).} This suggests that once the observed window becomes too small, it no longer provides enough spatial context to constrain the global interior structure. 
Even so, extremely limited observations are not entirely uninformative. Depending on the temporal stage of the flow, the observation window may fall partly or entirely in a quiescent region. In that case, the measured zero-velocity state still acts as a useful constraint by indicating where active motion is absent.

\textbf{Robustness against diluted spatial resolution.}
To evaluate the model's robustness to cases where measurements are only available on a sparse subset of tracers or pixels, which is relatively common in particle tracking velocimetry, this analysis let the boundary observation subject to spatial subsampling with a stride factor $k \in \{1, \dots, 9\}$. This reduces the available kinetic data to $1/k^2$ of the original resolution. 
As illustrated in Figure \ref{fig:ablation_combined} (right), the framework is relatively insensitive to this dilution for moderate levels of subsampling. The Pearson correlation $r$ remains close to 0.880 for stride factors up to $k=3$, meaning that the interior field can still be reconstructed well even when approximately 89\% of the measurement points are discarded (retaining only 11\% of the data). The performance begins to deteriorate at $k=4$, when the remaining observations become too sparse to represent the macroscopic velocity gradients along the boundary. 

\textcolor{black}{To confirm temporal representativeness, this analysis was further evaluated at the early acceleration ($t = 0.16$~s) and later deposition ($t = 0.76$~s) phases. As shown in Figure~\ref{fig:ablation_combined}, the error degradation trends and critical sparsity thresholds remain consistent across different macroscopic flow stages.} 
Overall, the sensitivity analysis shows that the proposed framework is reasonably robust to both truncated field of view and reduced spatial resolution. The main limitation is not moderate sparsity itself, but the loss of a sufficiently informative spatial anchor. This behavior is consistent with the intended use case: the model does not require dense boundary measurements, but it does require enough localized information to constrain the large-scale interior flow structure.

{\color{black}
\subsection{Limitations and opportunities}
While the proposed framework demonstrates the deployment of conditional flow matching to solve the limited observability challenges for inclined-plane granular flow, several limitations should be acknowledged and motivate further investigations:
\begin{enumerate}
    \item The CFM framework was trained based on a single geometry and material setup. Incorporating more geometric configurations and material properties would enhance the generalizability of the proposed framework.
    \item The velocity-to-physics decoder is a statistical regression rather than a constitutive surrogate. Implementing physical laws (e.g., mass and momentum conservation) into the AI framework would greatly enhance the mechanical robustness and model interpretability.
    \item The grid size is only twice the size of the DEM particle diameter. This provides a smoother look of field distributions but may cause numerical errors in grid-based physical variables since one particle may span multiple grid cells. This limitation can be mitigated by using larger grid size and by implementing physical laws like mass and momentum conservation.
    \item The grid-based stress calculation in Section 2.2 only considers the contact stress. While the inertial effects are negligible in the bulk, they may be significant in certain flow regions at certain time. These effects can be considered by incorporating full static and dynamic terms \cite{nicot2013definition} to better reflect the granular physics.
\end{enumerate}
}
\section{Conclusions}
This study presents \textcolor{black}{an innovative} application of conditional flow matching (CFM) to inverse reconstruction in granular flows. The proposed framework was developed to infer unobservable interior flow states from limited boundary measurements, a setting that is common in experiments but difficult to address with either conventional simulations or deterministic learning-based AI models. Starting from discrete element method (DEM) data, the particle-scale information was coarse-grained onto a fixed Eulerian grid to define cell-wise velocity and stress-related fields, and a conditional generative model was then trained to reconstruct interior velocity fields from boundary observations. A differentiable forward observation operator was used during sampling to enforce consistency with the measured boundary data, and a separate decoder was used to infer derived physical quantities, including mean stress $p$, deviatoric stress $q$, and granular temperature $T$. The results show that the framework provides a practical route for reconstructing hidden bulk states in dense granular flow while retaining the probabilistic character of the inverse problem. Main findings can be summarized as follows:

\begin{itemize}
    \item The framework accurately reconstructs interior velocity fields under both full and partial boundary observations, and further enables inference of mechanical and energy fluctuation fields such as $p$, $q$, and $T$. These results indicate that the framework can address the practical experimental problem of limited observability while extending the inference from kinematics to mechanics.
    \item Unlike deterministic models (e.g., CNN) that tend to return an over-smoothed average solution under incomplete observations, the CFM framework learns a distribution of admissible interior states conditioned on the limited measurements. This makes it better suited to ill-posed inverse problems in granular flow, where hidden regions cannot be inferred reliably through a single deterministic mapping.
    \item Another central contribution is that uncertainty is treated as part of the reconstruction itself rather than as a post-processing add-on. The framework identifies where the inferred interior state is well constrained and where ambiguity remains, which is essential for interpreting reconstructions from sparse measurements.
    \item \textcolor{black}{A major technical contribution of this work is the development of the sparsity-aware guidance mechanism. In granular systems characterized by extensive non-material empty spaces, standard error averaging artificially dilutes the correction gradient. The proposed unnormalized formulation avoids this dilution by} preserving the absolute physical scale of errors. This prevents unphysical predictions in non-material space, eliminates hyperparameter tuning, and ensures robust reconstructions under severely restricted experimental observations. The reconstruction can be reliably performed even with only 16\% of the observed area or when low-resolution observations only retain 11\% of the data.
\end{itemize}

Overall, this work introduces a new route for non-invasive, uncertainty-aware inverse inference of hidden bulk states in granular materials. More broadly, it shows how conditional generative modeling can connect sparse observations with interior kinematics and mechanics in particulate systems, and it opens the door to future developments of spatio-temporal reconstruction, forecasting, and tighter integration of physics-based constraints and richer target quantities.

\section*{CRediT authorship contribution statement}
\noindent
\textbf{Xuyang Li}: Conceptualization, Formal Analysis, Funding Acquisition, Investigation, Methodology, Software, Validation, Visualization, Writing – Original Draft, Writing – Review \& Editing.
\textbf{Rui Li}: Data Curation, Formal Analysis, Investigation, Software, Visualization.
\textbf{Teng Man}: Data Curation, Formal Analysis, Investigation, Software, Writing – Review \& Editing.
\textbf{Yimin Lu}: Conceptualization, Formal Analysis, Funding Acquisition, Investigation, Methodology, Project Administration, Supervision, Validation, Visualization, Writing – Original Draft, Writing – Review \& Editing.

\section*{Acknowledgments}
The authors acknowledge the Startup Funds from Texas Tech University (TTU) and University of North Carolina at Charlotte (UNC Charlotte), as well as the high performance computing resources from UNC Charlotte University Research Computing (URC), TTU High Performance Computing Center (HPCC), and the Texas Advanced Computing Center (TACC) from The University of Texas at Austin. The DEM simulations were conducted using MechSys, which is an open-access multiphysics simulation library, and its source code is available at \url{https://github.com/Axtal/mechsys.git}.

\providecommand{\noopsort}[1]{}\providecommand{\singleletter}[1]{#1}%

\clearpage
\appendix
\section{Additional Reconstruction Results}
While the main text details the velocity reconstruction at a representative time step, this appendix provides additional inference results at $t = 0.16$ s and $t = 0.36$ s to demonstrate the temporal consistency of the proposed framework.

\begin{figure}[H]
\centering
\includegraphics[width=0.7\textwidth]{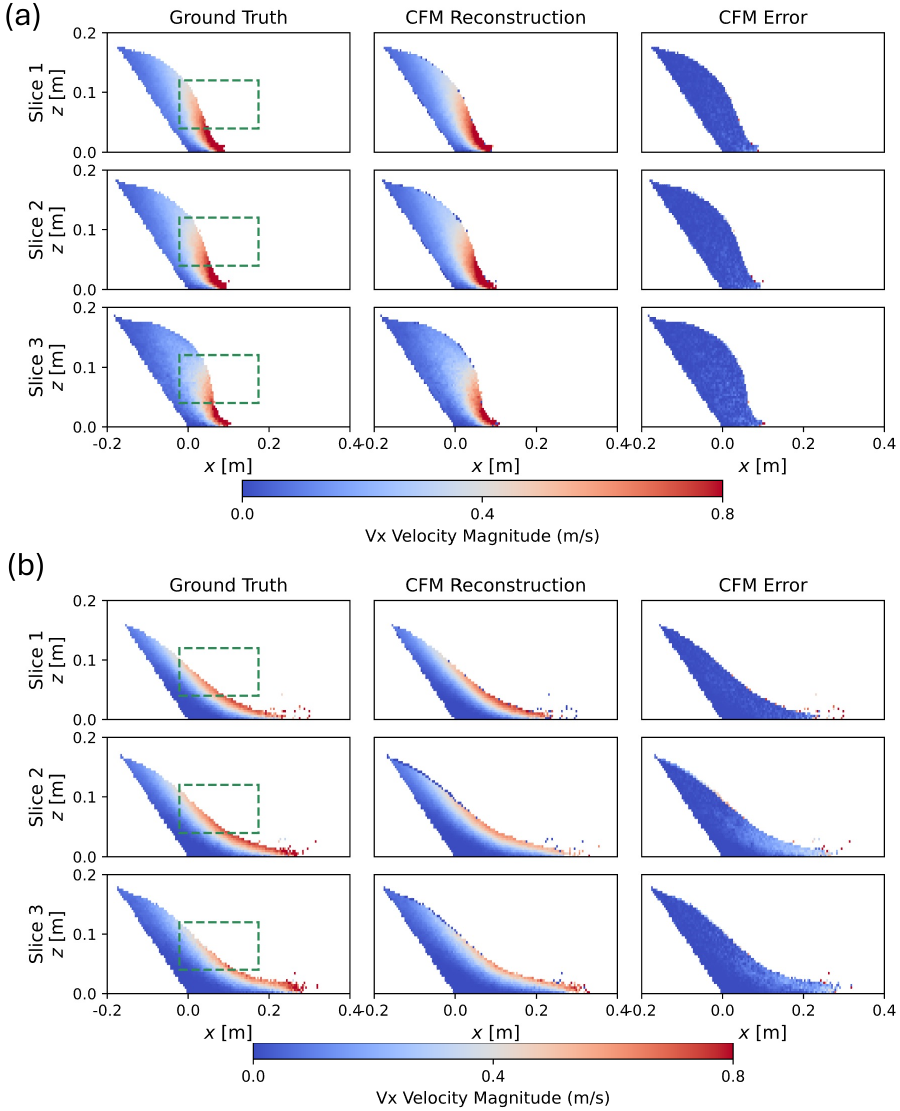}
\caption{Velocity field reconstruction under partial boundary observations at additional time steps. (a) Reconstructed slices at $t = 0.16$ s. (b) Reconstructed slices at $t = 0.36$ s. Both panels compare the DEM ground truth, the generative prediction, and the absolute spatial error across three distinct internal depths.}
\label{fig:appendix1}
\end{figure}

Figure \ref{fig:appendix1} illustrates the reconstructed velocity fields for three internal slices at these two time steps, following the same partial observation setup and 3x3 layout used in the main text. The quantitative metrics confirm stable performance across different dynamic stages of the granular flow. When evaluated strictly within the active flowing regions, the spatial correlation ($r_x$) consistently remains above 0.85, and the active RMSE maintains a low magnitude for both the initial acceleration phase ($t = 0.16$ s) and the developed flow stage ($t = 0.36$ s). These results verify that the generative prior is robust and applicable throughout the entire temporal evolution of the granular flow.

\begin{figure}[H]
\centering
\includegraphics[width=\textwidth]{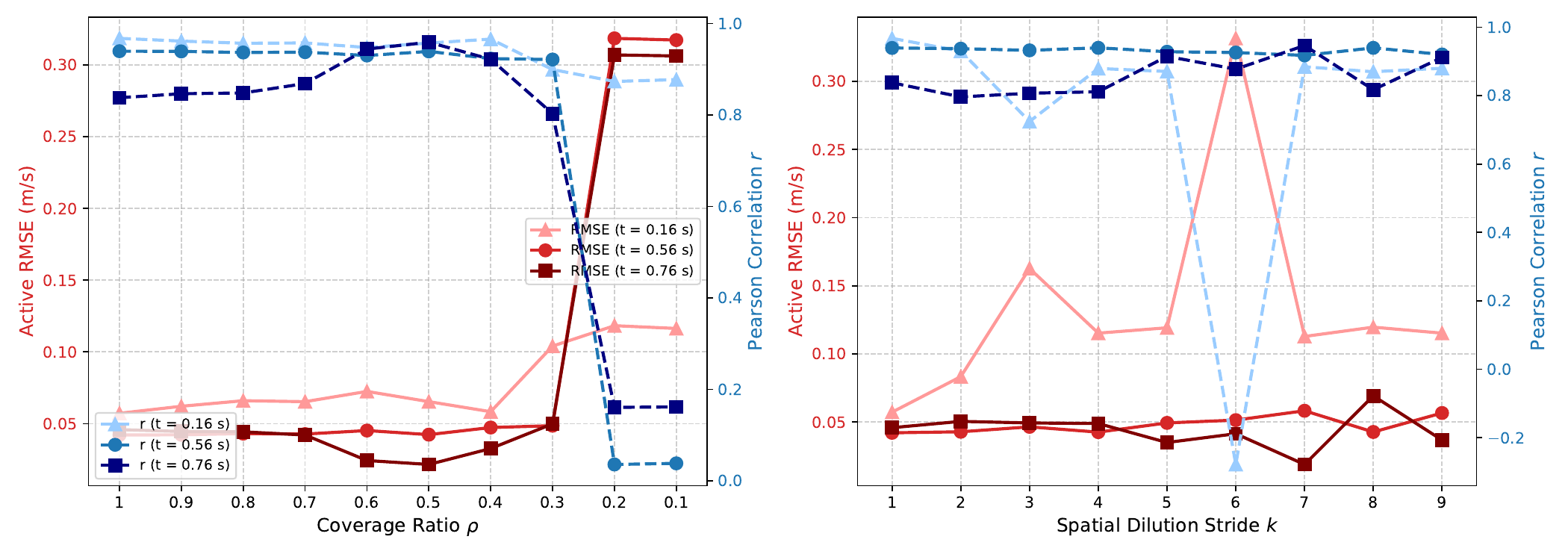}
\caption{Sensitivity analysis of the reconstruction fidelity evaluated on Slice 2 across three different macroscopic flow stages ($t = 0.16, 0.56,$ and $0.76$~s). The relative degradation patterns remain consistent with the deepest slice.}
\label{fig:ablation_s6}
\end{figure}

\begin{figure}[H]
\centering
\includegraphics[width=\textwidth]{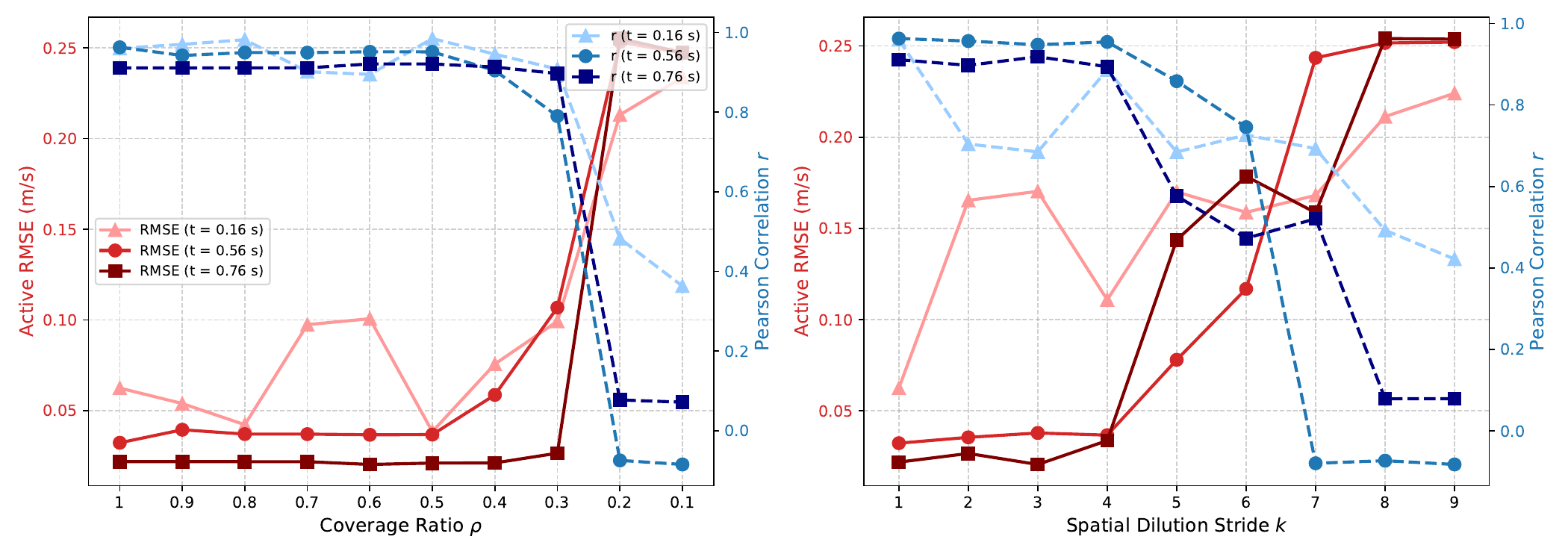}
\caption{Sensitivity analysis of the reconstruction fidelity evaluated on Slice 1 across three different macroscopic flow stages ($t = 0.16, 0.56,$ and $0.76$~s). This slice is closest to the observation boundary, yielding the lowest absolute reconstruction errors while maintaining the same robustness threshold.}
\label{fig:ablation_s3}
\end{figure}

\textcolor{black}{
\section{Additional Ablation Results}\label{appendix2}
Following the discussion in Section \ref{sec:ablation_obs}, we provide additional sensitivity evaluations for the shallower interior slices. Figure \ref{fig:ablation_s6} and Figure \ref{fig:ablation_s3} present the reconstruction performance under varying observation constraints for Slice 2 (middle depth) and Slice 1 (shallow depth), respectively. Consistent with the deepest slice analyzed in the main text, these shallower planes exhibit identical robustness trends. Expectedly, they yield overall lower absolute RMSE values due to their closer proximity to the boundary observations.}

\end{document}